\documentclass[a4paper,english]{article}
\usepackage{amsmath,amsfonts,amssymb,amsthm}
\theoremstyle{definition}

\usepackage{authblk}
\usepackage{caption}
\usepackage{color}
\usepackage[latin1]{inputenc}
\usepackage[autostyle]{csquotes}
\usepackage{dsfont}
\usepackage{fontawesome}
\usepackage[multiple]{footmisc}
\usepackage{geometry}
\usepackage{graphicx}
\usepackage{hyperref}
\usepackage{listings}
\usepackage{multicol}
\usepackage{nameref}
\usepackage{natbib} 
\usepackage{setspace}
\usepackage{titlesec}
\usepackage{textcomp} % for apostrophes within a quotation ex: `It\textquotesingle s a nice day!'
\usepackage{wallpaper}%
\usepackage{xspace}

\usepackage{babel}
\usepackage{mathpazo}
\usepackage[T1]{fontenc}

\def\title#1{{\Large\bf  \begin{center} #1 \vspace{0pt} \end{center}  } }
\def\authors#1{{ \begin{center} #1 \vspace{0pt} \end{center} } }
\def\inst#1{\unskip$^{#1}$}

\newcommand{\keywords}[1]{\bigskip \noindent {\bf Keywords:} \ #1}

\makeatletter
\newcommand*\bigcdot{\mathpalette\bigcdot@{.5}}
\newcommand*\bigcdot@[2]{\mathbin{\vcenter{\hbox{\scalebox{#2}{$\m@th#1\bullet$}}}}}
\makeatother

\definecolor{gray}{rgb}{0.95,0.95,0.95}
\titleformat{\section}
  {\normalfont\fontsize{12}{15}\bfseries}{\thesection}{1em}{}
  
  \titleformat{\subsection}
  {\normalfont\fontsize{11}{15}\bfseries}{\thesubsection}{1em}{}

  \titleformat{\subsubsection}
  {\itshape\fontsize{11}{15}}{\thesubsection}{1em}{}
 
\makeatletter
\renewenvironment{abstract}{%
    \if@twocolumn
      \section*{\abstractname}%
    \else %% <- here I've removed \small
      \begin{center}%
        {\bfseries \large\abstractname\vspace{\z@}}%  %% <- here I've added \large
      \end{center}%
      \quotation
    \fi}
    {\if@twocolumn\else\endquotation\fi}
\makeatother

\providecommand{\keywords}[1]
{
  \small	
  \textbf{\textit{Keywords---}} #1
}

\begin{document}

%\textsc{\newcommand{\Title}{Title}}%
%\rightlogo[1]{Picture1}
%\newcommand{\Author}{\normalsize Rui J. Costa, Moritz Gerstung}%
%\author{\Author}
%\newcommand{\Affiliation}{$^1$European Bioinformatics Institute (EMBL-EBI), Hinxton, UK} %
%\affiliation{\Affiliation}
%\newcommand{\Email}{rui.costa@ebi.ac.uk}%
%\email{\Email}}

% 1. Fill in the abstract title
\title{The R package \texttt{ebmstate} for disease progression analysis under empirical Bayes Cox models}

% 2. Fill in all the authors, the presenting author must be first author 
\authors{
%  Rui J. Costa\inst{1} and
%  Moritz Gerstung\inst{1} 
Rui J.  Costa \inst{1} and Moritz Gerstung \footnote{\label{note1} European Bioinformatics Institute - EMBL-EBI,  Hinxton,  CB10 1SD, United Kingdom}\inst{,}\footnote{\label{note2} Genome Biology Unit, EMBL, Meyerhofstrasse 1, 69117 Heidelberg, Germany}\inst{,}\footnote{\label{note3} German Cancer Research Center (DKFZ), Im Neuenheimer Feld 280, 69120 Heidelberg, Germany\\}
}

% 3. Fill in all the affiliations. Include city and country of each institute, do not include the full address.
%\university{
%  \inst{1} European Bioinformatics Institute (EMBL-EBI)\\
%}

%%% Begin of Multicols-Enviroment
\vspace{0.6cm}

\begin{abstract}
\small
Statistical inference about multi-state random processes relies often on models defined by transition hazard rates. Inference based on such models necessarily involves the estimation of these rates, but it can also include estimates of state occupation probabilities. If the multi-state model incorporates a regression model, estimates of hazard rates and state occupation probabilities can be personalised, i.e. they can refer to an individual with specific covariate measurements. 
The software package \texttt{mstate}, in articulation with the package \texttt{survival}, provides not only a well-established multi-state survival analysis framework in R, but also one of the most complete, as it includes point and interval estimation of relative transition hazards, cumulative transition hazards and state occupation probabilities, both under clock-forward and clock-reset models; personalised estimates can also be obtained by fitting a Cox regression model.
The new R package \texttt{ebmstate}, which we present in the current paper, is an extension of \texttt{mstate} and, to our knowledge, the first R package for multi-state model estimation that is suitable for higher-dimensional data and complete in the sense just mentioned.  Its extension of \texttt{mstate} is threefold: it transforms the Cox model into a regularised, empirical Bayes model that performs significantly better with higher-dimensional data; it replaces asymptotic confidence intervals meant for the low-dimensional setting by non-parametric bootstrap confidence intervals; and it introduces an analytical, Fourier transform-based estimator of state occupation probabilities for clock-reset models that is substantially faster than the corresponding, simulation-based estimator in \texttt{mstate}.  
The present paper includes a detailed tutorial on how to use our package to estimate transition hazards and state occupation probabilities, as well as a simulation study showing how it improves the performance of \texttt{mstate}.
\end{abstract}


%
%\blfootnote{
%\noindent \faEnvelopeO \hspace{0.3cm}  Rui J.  Costa
%\par \hspace{0.95cm} rui.costa@ebi.ac.uk}
\section{INTRODUCTION}

Multi-state models based on transition hazard functions are often used in the statistical analysis of longitudinal data, in particular disease progression data \citep{Hougaard1999}. The multi-state model framework is particularly suitable to accommodate the growing level of detail of modern clinical data: as long as a clinical history can be framed as a random process which, at any moment in time, occupies one of a few states, a multi-state model is applicable. Another strong point of this framework is that it can incorporate a \textit{regression model}, i.e., a set of assumptions on how covariates, possibly time-dependent ones, affect the risk of transitioning between any two states of the disease. Once estimated, multi-state models with regression features allow the stratification of patients according to their transition hazards. In addition, it is possible, under some models, to generate disease outcome predictions. 
These come in the form of \textit{state occupation probability} estimates, meaning estimates of the probability of being in each state of the disease over a given time frame.

The survival analysis `task view' of the Comprehensive R Archive Network lists six R packages that are able to fit \textit{general} multi-state models and, at the same time, feature some kind of regression model or algorithm: \texttt{msm} \citep{Jackson2011}, \texttt{SemiMarkov} \citep{Listwon2015}, \texttt{survival} \citep{survival_package}, \texttt{mstate} \citep{Wreede2010}, \texttt{mboost} \citep{mboost_package} -- as extended by \texttt{gamboostMSM} \citep{gamboostMSM_package} -- and  \texttt{penMSM} \citep{penMSM_package}. All of them implement relative risk regression models, in particular the Cox model. The only exception is \texttt{survival}, which fits both the Cox model and Aalen's additive regression model \citep{Aalen1989}. 

Recall that in a semi-parametric Cox regression model each transition hazard is assumed to be the product of a baseline hazard function of unspecified form (the non-parametric component) and an exponential relative hazard function (the parametric component) \citep[][p. 133]{Aalen2008}. 
The Cox models implemented in these packages are all semi-parametric, with the exception of those in \texttt{msm} and \texttt{SemiMarkov}, which are fully parametric, i.e. they also restrict the baseline hazards to specific parametric families. In \texttt{msm} and \texttt{SemiMarkov}, the stronger assumptions regarding the functional form of the hazard are leveraged to do away with other common assumptions: \texttt{SemiMarkov} drops the usual Markov property to implement homogeneous semi-Markov models; \texttt{msm} is suitable for \textit{panel data}, i.e., data in which the state of each individual is known only at a finite series of times. 

Packages \texttt{penMSM} and \texttt{gamboostMSM} are the best suited to deal with higher-dimensional covariate data. 
The first of these packages relies on a structured fusion lasso method, while the second implements (jointly with \texttt{mboost}) a boosting algorithm. Both methods induce sparsity in the number of non-zero covariate effects, as well as equality among the different transition effects of each covariate, and are thus especially useful to reduce complicated multi-state models to more interpretable ones. The remaining packages  assume the standard, fixed effects Cox model and do not include regularisation or variable selection features.

It is also illustrative to order the six packages mentioned according to how extensive their analysis workflow is. Packages \texttt{SemiMarkov} and \texttt{penMSM} are intended for the estimation of relative transition hazards  only (i.e., for estimating the impact of covariates on each transition hazard). With the package \texttt{mboost} (as extended by \texttt{gamboostMSM}) it is also possible to estimate the baseline transition hazards. Finally, a more complete workflow including estimates of both relative and cumulative transition hazards, as well as state occupation probabilities, is implemented in \texttt{msm} and \texttt{mstate}, and has been under implementation in \texttt{survival} (version 3.0 or later).

The present paper provides an introduction to \texttt{ebmstate}, a new R package for multi-state survival analysis available for download on the Comprehensive R Archive Network (CRAN).
The main goal of \texttt{ebmstate} is to provide an analysis framework for the Cox model that performs better with higher-dimensional covariate data and is also complete, in the sense of being able to generate point and interval estimates of relative transition hazards, cumulative transition hazards and state occupation probabilities, both under clock-forward and clock-reset models. 
 A fundamental characteristic of \texttt{ebmstate} is that it re-implements and extends the analysis framework of \texttt{mstate},  which is complete in the sense just mentioned.  In fact,  to a large extent,  our package was built by importing, adapting and replacing functions from the \texttt{mstate} package. This not only eliminates redundancies, but also makes our package more accessible to the numerous users of \texttt{mstate} (the three papers associated with \texttt{mstate} have jointly over 2000 citations). 
 
To improve the performance of \texttt{mstate}'s multi-state Cox model when dealing with higher-dimensional covariate data, a ridge-type regularisation feature was added.  We allow the  regression coefficients of the model to be partitioned into groups, with each group having its own Gaussian prior. A group can gather, for example, all the regression coefficients for a given transition. Or, within a given transition,  coefficients can be grouped according to the covariate type they refer to  (for example, demographic, clinical or genomic type).
  The resulting hierarchical Bayes model is \textit{empirical} in that a full prior elicitation is not required (the mean and variance hyper-parameters of the Gaussian are estimated from the data). Model fitting relies on the iterative algorithm introduced by \citet{Schall1991}, which typically converges after a small number of steps. A simulation study  showing that Schall's algorithm performance compares well with that of other algorithms for ridge penalty optimisation, including one based on cross-validation, can be found in \citet{Perperoglou2014}.

The asymptotic confidence intervals generated by \texttt{mstate} are applicable when the number of observations is much larger than the number of parameters to be estimated (see section \ref{sec:interval_estimation} below).  
To preserve the completeness of \texttt{mstate}'s framework in higher-dimensional settings, we therefore implemented non-parametric bootstrap intervals of regression coefficients, cumulative transition hazards and state occupation probabilities.  

\begin{figure}[h] 
\centering         
\includegraphics[width=14.5cm, angle=0]{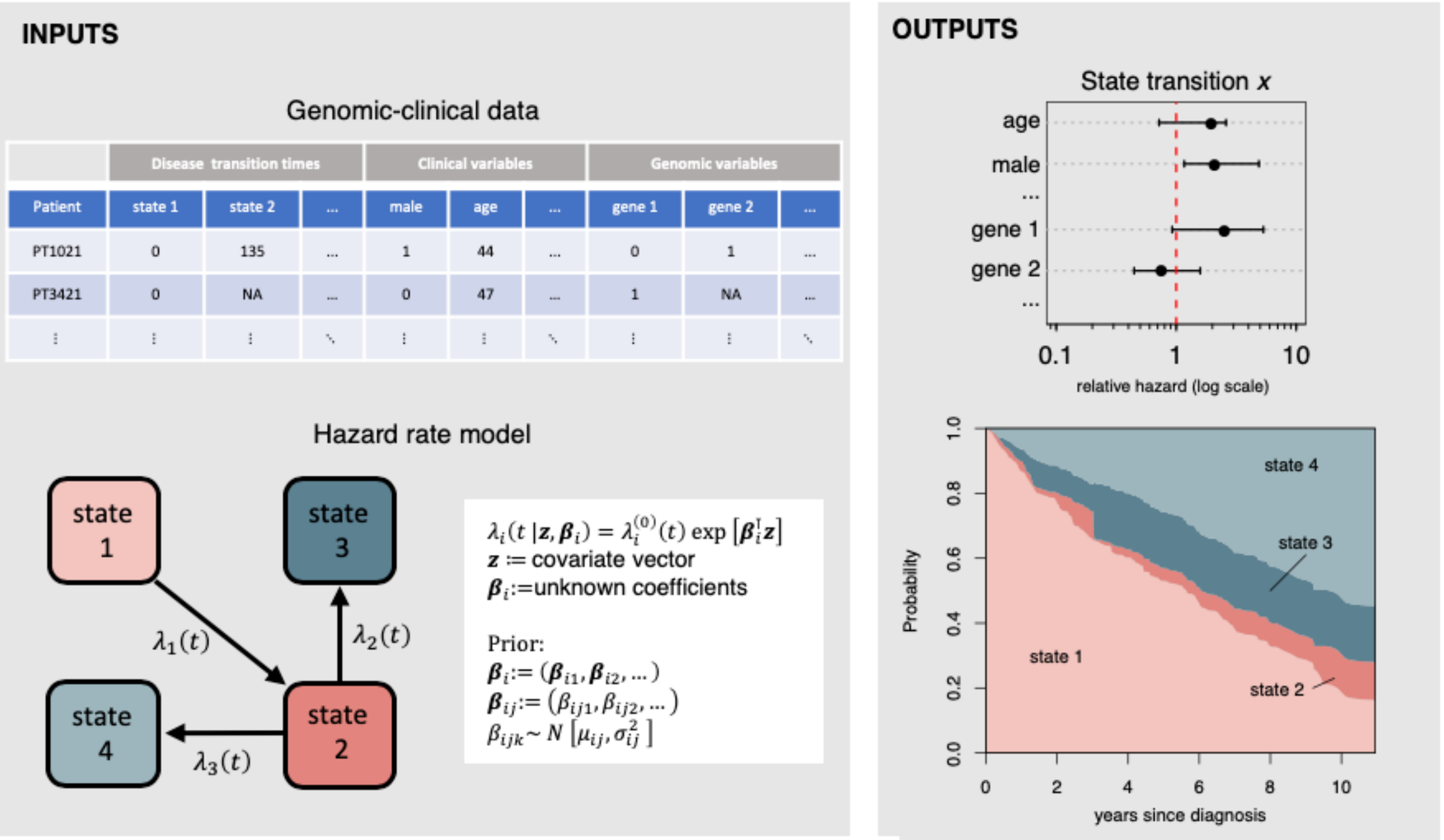} 
\caption{Summary of inputs and outputs of the package \texttt{ebmstate}.  The input data set should be one that violates the assumption -- commonly used in survival analysis -- that the number of observations is much larger than the number of parameters to be estimated (a genomic-clinical data set is shown as a typical example). The input model is a multi-state Cox model defined by a transition structure and a prior distribution on the regression coefficients.  This prior distribution is defined by partitioning the vector of regression coefficients into groups of regression coefficients, with each group having its own Gaussian prior with undetermined mean and variance.  The outputs of \texttt{ebmstate} include estimates of the relative transition hazards associated with each covariate, as well as estimates of the probability that a specific patient (with specific covariate measurements) has of occupying each state of the model over some time period. Estimates of cumulative transition hazards are omitted from the figure.}      
\label{fig:package_summary_figure} 
\end{figure}

The high computational cost implied by the non-parametric bootstrap motivated a third extension to \texttt{mstate}. We developed an estimator of state occupation probabilities under clock-reset Cox models  that is based on a convolution argument \citep[as in][]{Spitoni2012} and the Fast Fourier transform (FFT).  At present,  the estimation of such probabilities for clock-forward Cox models can be carried out using the efficient,  product-limit based algorithm available in \texttt{mstate}.  However, for clock-reset Cox models,  only a simulation-based estimator is available in this package.  The FFT estimator in \texttt{ebmstate} was conceived as a faster alternative to this simulation-based estimator,  but its scope is currently restricted to multi-state models with a tree-like transition structure, i.e.  in which a transition between two states is either not possible or follows a unique sequence of states.
 Figure \ref{fig:package_summary_figure} provides a short graphical summary of \texttt{ebmstate}, with the main inputs -- a genomic-clinical data set and an empirical Bayes multi-state Cox model -- and the main outputs -- the estimates of relative hazards and state occupation probabilities (cumulative transition hazards are omitted).
 
 As already mentioned, our empirical Bayes method improves estimator performance in models with larger numbers of covariates (see section \ref{sec:estimator_performance} on estimator performance). 
Also, as a ridge-type regression method, it can be used as an alternative to the lasso method of \texttt{penMSM} in two particular cases:  when the levels of correlation between covariates are high enough to compromise the stability of lasso-based covariate selection; or simply to improve prediction accuracy when interpretability is not essential and the number of covariates is not greater than the number of observations \citep{Zou2005}.
In addition, and perhaps more importantly,  \texttt{ebmstate} goes beyond the  regularised estimation of transition hazards offered by \texttt{penMSM} and \texttt{gamboostMSM}: point and interval estimates of state occupation probabilities under the regularised Cox model can also be computed.

\section{MODELS}

A multi-state Cox model is a continuous-time stochastic process with a finite (and usually small) state space $\mathcal{S}$. 
 To better describe the models implemented in \texttt{ebmstate}, we define the following notation. We let $t$ denote the time since some initiating event (usually diagnosis or disease onset). For $t \in \left[0, \infty\right)$, we define the following random variables: $X(t)$ represents the disease state of the patient, $S(t)$ the time spent in the current state,  and $\vec{Z}\left(t\right)$  the value of a covariate vector.  
The realisation of each component of the process $\lbrace\vec{Z}\left(t\right)\rbrace$ is a step function, possibly approximating the evolution in time of a continuous covariate.
 In addition, $\lbrace\vec{Z}\left(t\right)\rbrace$ is assumed not-adapted to the filtration generated by $\lbrace X\left(t\right)\rbrace$ (an adapted covariate is one whose path until $t$ is known once $\lbrace X \left(u\right)\rbrace$, $u \leq t$, is known). 
 The transition hazard rate of a patient from state $i$ to state $j$ ($i\neq j$) at time $t$, conditional on the sojourn time and the covariate vector, is defined as
 \begin{align*}
 &\alpha_{ij}\left(t|\mathbf{z},s \right):=\lim_{h \downarrow 0}\frac{1}{h}\mathrm{P}\left[X(t+h)=j\,|\,X(t)=i,S(t)=s,\vec{Z}(t)=\mathbf{z} \right]\;, \;s\in \left[0,\infty\right)\;,\;t\in \left[s,\infty\right)\;.
\end{align*}
Independent right-censoring and left-truncation are assumed throughout \citep[][p. 57]{Aalen2008}.  The purpose of the present section is to give a (not necessarily exhaustive) description of the scope of \texttt{mstate} and \texttt{ebmstate} with respect to the multi-state Cox model.  Using the terminology in \citet{Putter2011},  a Cox model is termed a `clock-reset' model when
\begin{align}
\label{eq:clock_reset_Cox}
\alpha_{ij}\left(t\,|\,\mathbf{z}, s\right)&=\lambda_{ij}^{(0)}\left(s\right)\exp\left[ \boldsymbol{\beta}^{\intercal}_{ij}\,\mathbf{z}\right] \quad,
\end{align}
and it is termed a `clock-forward' model when
\begin{align}
\label{eq:clock_forward_Cox}
\alpha_{ij}\left(t\,|\,\mathbf{z}\right)&=\alpha_{ij}^{(0)}\left(t\right)\exp\left[ \boldsymbol{\beta}^{\intercal}_{ij}\,\mathbf{z}\right] \quad.
\end{align}
In both cases, $i,j \in \mathcal{S}$, with $i\neq j$; $\boldsymbol{\beta}_{\scriptscriptstyle ij}$ is an unknown vector of regression coefficient parameters, and both $\lambda^{\scriptscriptstyle (0)}_{ij}(\cdot)$ and $\alpha^{\scriptscriptstyle (0)}_{ij}(\cdot)$ are unknown (baseline hazard) functions, non-negative on $\mathds{R}^{+}$.  When, as in equation \ref{eq:clock_reset_Cox},  $\alpha_{ij}\left(t|\mathbf{z},s\right)$ is the same for all $t\geq s$, we simplify its notation to $\lambda_{ij}\left(s|\mathbf{z}\right)$.  As can be seen from equations \ref{eq:clock_reset_Cox} and \ref{eq:clock_forward_Cox},  the `clock-reset' and `clock-forward' models are models for how the transition hazard rates are affected by time. In the former case, the only relevant time scale is the time $s$ spent in the current state, whereas in the latter only the time $t$ since the initiating event matters. 
While the `clock-forward' model is arguably the default one in multi-state survival analysis \citep{Andersen1993,Aalen2008}, in some cases the `clock-reset' model is more appropriate.  For example,  in some forms of cancer,  it can be sensible to assume that the transition hazards from the state of complete remission depend on the sojourn time, rather than on the time since the initial diagnosis.

\subsection{Relative transition hazards}
\label{sec:models_relative_hazards}
The parametric component of the transition hazard from $i$ to $j$, written $\exp\left[\boldsymbol{\beta}^{\intercal}_{ij} \,\mathbf{z}\right]$, is termed the relative transition hazard.  In \texttt{mstate} and \texttt{ebmstate},  estimating the relative transition hazard amounts to estimating the regression coefficient vector $\boldsymbol{\beta}_{ij}\,$.  
In \texttt{mstate}, these parameters are assumed to be non-random. With \texttt{ebmstate}, the following prior distributions  can be imposed.

Define $\mathcal{P}$ as the set of all pairs of states between which a direct transition is possible. Let $\lbrace \boldsymbol{\beta}_{\scriptscriptstyle ij} \rbrace $, for all $(i, j) \in \mathcal{P}$, be a partition of $\boldsymbol \beta$, a vector containing the regression coefficients for all direct transitions allowed. Each $\boldsymbol{\beta}_{\scriptscriptstyle ij}$ is further partitioned into  $\lbrace \boldsymbol{\beta}_{\scriptscriptstyle ijk} \rbrace$, for $k \in \left\lbrace 1,2,...,n_{\scriptscriptstyle ij} \right\rbrace$. In \texttt{ebmstate}, the most general model regarding the prior distribution  of $\boldsymbol{\beta}$ makes two assumptions:  a) the scalar components of $\boldsymbol{\beta}$ are independent and normally distributed; b) the scalar components of $\boldsymbol{\beta}_{\scriptscriptstyle i j k}$ have a common (and undetermined) mean $\mu_{\scriptscriptstyle ijk}$ and a common (and also undetermined) variance $\sigma^{2}_{\scriptscriptstyle ijk}\;$.

The purpose of the framework just described is to allow the clustering of covariate effects according to their prior distribution.
If there is no prior knowledge about how this clustering should be done,  a single Gaussian prior can be imposed on all regression coefficients at once. If prior knowledge allows the grouping of effects according to the transition they refer to, a different Gaussian prior can be assigned to the coefficients of each transition. Even within each transition,  different groups of coefficients  can be assigned different prior distributions. In the analysis of biomedical data, for example, there can be a split between genes which are known to affect the transition hazard, and other genes whose effect is unknown.

\subsection{Cumulative transition hazard functions}

Our package imports from \texttt{mstate} a Breslow estimator of two types of cumulative transition hazard: one on a global time scale,  defined as
\begin{align*}
\mathrm{A}_{ij}\left(t\,|\,\mathbf{z}\right)&:=\int_{0}^{t}\alpha_{ij}^{(0)}\left(u\right)\exp\left[ \boldsymbol{\beta}^{\intercal}_{ij}\,\mathbf{z}\right]\mathrm{d}u\quad,
\end{align*}
and another on a sojourn time scale, defined as
\begin{align*}
&\Lambda_{ij}(s\,|\,\mathbf{z}):=\int_{0}^{s}\lambda_{ij}^{(0)}\left(u\right)\exp\left[ \boldsymbol{\beta}^{\intercal}_{ij}\,\mathbf{z}\right]\mathrm{d}u\quad.
\end{align*}
Note that, in either case, the covariate vector is assumed to remain constant.

\subsection{State occupation probabilities}
By state occupation probability, we mean the probability that a patient in state $i$ at time $0$ finds herself in state $j$ at time $t$.  The estimates of these probabilities can be seen as functionals of the estimated cumulative transition hazard functions.  For this reason,  the restriction to models with time-fixed covariates,  which was just seen to be applicable to the estimators of cumulative transition hazards,  carries over to the estimation of state occupation probabilities.  

When conditioning on a given covariate path (time-fixed or not), state occupation probability estimates are not valid unless the covariates are \textit{external} \citep[][p. 142]{Cortese2010,Aalen2008}. 
Note that a vector of covariates $\lbrace \vec{Z}(u)\rbrace_{u\geq 0}$ is said to be \textit{external} if, for all $t \in \left[0,\infty\right)$, each transition hazard at $t$, conditional on $ \vec{Z}(t)$, is independent of $\lbrace \vec{Z}(u)\rbrace_{u>t}$ (i.e.  independent of the future path of the covariate). Otherwise, it is said to be \textit{internal} \citep[for more details on the distinction between internal and external covariates, see][chapter 6]{Kalbfleisch2002}.
When one does not wish (or is not possible due to $\vec{Z}$ being \textit{internal}) to condition on a future  covariate path of the covariate process, the uncertainty introduced by  this process needs to be accounted for.  This can be done by extending the state space of the disease process, so that it includes information on the disease \textit{and} the covariate process \citep[][p. 170]{Andersen1993}.  For example, to include a dichotomous transplant covariate (an internal covariate) in a simple survival model with two states, the state space is expanded from $\lbrace$alive,  deceased$\rbrace$ to $\lbrace$alive without transplant, alive with transplant,  deceased$\rbrace$. One can then either assume that transplanted patients have a different baseline death hazard or, more simply,  that transplantation scales the death hazard by some constant $\exp \left( \gamma\right)$.  A similar but more detailed example can be found in \citet[][section 2.3.2, `model 3' ]{Wreede2010}.

\section{ESTIMATION}
In the current section,  we present the estimation methods underlying the extensions of \texttt{mstate} implemented in \texttt{ebmstate}.  
 \label{sec:estimation}

\subsection{Relative and cumulative hazard functions}
Let $\boldsymbol{\mu}_{\scriptscriptstyle ij}$, with $\left(i,j\right) \in \mathcal{P}$ (the set of direct transitions allowed), denote a vector whose scalar components are the parameters $\mu_{\scriptscriptstyle ijk}$,  $k \in \left\lbrace 1,2,...,n_{\scriptscriptstyle ij} \right\rbrace$.  Similarly, let $\boldsymbol{\sigma}^{2}_{\scriptscriptstyle ij}$ be composed of the parameters  $\left\lbrace \sigma^{2}_{\scriptscriptstyle ijk}\right\rbrace_{k}$. The estimation of  $\boldsymbol{\beta}$, $\boldsymbol{\mu}:=\lbrace\boldsymbol{\mu}_{\scriptscriptstyle{ij}}\rbrace$  and  $\boldsymbol{\sigma}^2:=\lbrace\boldsymbol{\sigma}^2_{\scriptscriptstyle ij }\rbrace$ relies on the restricted maximum-likelihood (REML) type algorithm described in \cite{Perperoglou2014}, and introduced by \cite{Schall1991}. The resulting estimate of $\boldsymbol{\beta}$ is a maximum \textit{a posteriori} estimate; the estimates of $\boldsymbol{\mu}$ and $\boldsymbol{\sigma}^{2}$ are empirical Bayes estimates. In \texttt{ebmstate}, the estimator based on this algorithm  is implemented in the function \texttt{CoxRFX} .  The results of a simulation study showing its consistency are included in the Supplementary Materials (file ESM\_1.html, section 1).

The computation of cumulative hazard rates for given covariate values  and an estimated regression coefficient vector relies on the function \texttt{msfit\_generic}, which is essentially a wrapper for the function \texttt{mstate::msfit} (see section \ref{sec:computing_cumulative_hazards}). For the mathematical details of this computation, we refer therefore the reader to \citet{Wreede2010}.

\subsection{State occupation probabilities}
\label{sec:trans_probs}
The package \texttt{mstate} includes a simulation-based estimator that can take as input either  $\hat{\mathrm{A}}_{ij}\left(\cdot\,|\,\mathbf{z}\right)$ or $\hat{\Lambda}_{ij}\left(\cdot\,|\,\mathbf{z}\right)$ to generate estimates of state occupation probabilities under the clock-forward or the clock-reset model respectively.  
Another available estimator,  an Aalen-Johansen-type estimator based on product integration, is far more efficient computationally and takes as input $\hat{\mathrm{A}}_{ij}\left(\cdot\,|\,\mathbf{z}\right)$ only.  As the scope of this estimator has been restricted to clock-forward Cox models \citep{Andersen1993,Aalen2008}, in our package we implemented a convolution-based estimator as a computationally efficient alternative (for models with a tree-like transition structure).  
\subsubsection*{Convolution-based methods for clock-reset Cox models}
\label{sec:semiMarkov}

For convenience, let the sequence of states from $0$ to $n$ have the labels $0,1,2,...,n\,$, where $0$ is the initial state by definition, and $n$ is some state that might (eventually) be reached by the process. In addition, define $X_{0}:=X(0)$ and $T_{0}:=0$, and let $\left(X_{i},T_{i}\right)$, $i \in \left\lbrace 1,2,... \right\rbrace$, denote the marked point process associated with $\left\lbrace X(t)\right\rbrace$, so that $T_{i}$ is the time of the $i^{th}$ transition and $X_{i}$ is the state the process jumps to at time $T_{i}$. 
The inter-transition times are denoted by  $\tau_{ij}:=T_{j}-T_{i}$, for $j>i$.
We can write the probability that a patient in state $0$ at time $0$ finds herself in state $n$ at time $t$, conditional on $\vec{Z}(u)=\mathbf{z}$ for all $u \geq 0$,  as
\begin{align*}
 &\mathrm{P}\left[X(t)=n\,|\,X(0)=0\,, \vec{Z}(u)=\mathbf{z},\,u \geq 0 \right]\\
 &\,=\mathrm{P}\left[X_{n}=n,\tau_{0,n} < t,\tau_{n,n+1}\geq t- \tau_{0,n} |X_{0}=0\,,  \vec{Z}(u)=\mathbf{z},\,u \geq 0 \right] \,.\nonumber
\end{align*}

Recall that $\lambda_{i,i+1}\left(s\,|\, \mathbf{z}\right)$ denotes the hazard rate of a transition to state $i+1$ at time $s$ since arrival in state $i$, for a  patient that has covariate vector $\mathbf{z}$. The cumulative hazard for the same transition between sojourn times $0$ and $s$, if the patient's covariate vector remains constant at $\mathbf{z}$, is represented by $\Lambda_{i,i+1}\left(s \,|\, \mathbf{z}\right):=\int_{0}^{s}\lambda_{i,i+1}\left(x\,|\, \mathbf{z}\right)\mathrm{d}x$.
 Similarly, we let $\lambda_{i}\left(s\,|\, \mathbf{z}\right)$ represent the hazard rate of going to any state that can be reached directly from $i$, at time $s$ since arrival in state $i$, for a patient with covariate vector $\mathbf{z}$. The cumulative hazard for the same event between sojourn times $0$ and $s$, if the patient's covariate vector remains constant at $\mathbf{z}$, is represented by $\Lambda_{i}\left(s \,|\, \mathbf{z}\right)$.
 The expressions $\hat{\Lambda}_{i}\left(s \,|\, \mathbf{z}\right)$ and $\hat{\Lambda}_{i,i+1}\left(s \,|\, \mathbf{z}\right)$ denote the Breslow estimators of the cumulative hazards just defined.
 In what follows, all references to probabilities,  hazard rates and cumulative hazards are to be understood as conditional on  $\vec{Z}(u)=\mathbf{z}\,$, for $u\geq 0$: this condition is omitted to simplify the notation. 

 In \texttt{ebmstate}, the function \texttt{probtrans\_ebmstate} generates a set of state occupation probability estimates at equally spaced time points:
\begin{align*}
&\left\lbrace \hat{p}_{0n}\left(k\right)\right\rbrace_{k} :=\left\lbrace \hat{\mathrm{P}}\left[X_{n}=n,\tau_{0,n} < t_{k},\tau_{n,n+1}\geq t_{k}- \tau_{0,n}\,|\, X_{0}=0 \right] \right\rbrace_{k}\;,\; k=0,1,2,...,K\,;\, t_{k}=k\times \Delta t \;.
\end{align*}
The number $K$ of time intervals is $10,000$ by default and $t_{K}$ is a parameter set by the user.
 Defining the functions
\begin{align*}
q_{ij}\left(k\right):=\mathrm{P}\left[X_{j}=j, \tau_{ij}\in \left[t_{k},t_{k+1}\right)\,|\,X_{i}=i\right]
\end{align*} 
and 
 \begin{align*}
r_{i}\left(k\right):=\mathrm{P}\left[\tau_{i,i+1} > t_{k} \,|\,X_{i}=i\right]\;,
\end{align*} 
the algorithm behind \texttt{probtrans\_ebmstate} can be described as follows:
\begin{enumerate}
\item For $j=1,2,...,n$,  compute
\begin{flalign}
\label{eq:est1}
 \hat{q}_{j-1,j}\left(k\right)&:=\exp \left[-\hat{\Lambda}_{j-1}\left(t_{k}\right)\right]\Delta \hat{\Lambda}_{j-1,j}\left(t_{k}\right)&&
\end{flalign}
 for $k=0,1,...,K-1$.
\item For $j=2,3,...,n$, compute (iteratively)
\begin{flalign}
\label{eq:est2}
 \hat{q}_{0j}\left(k\right):=&\sum_{l=0}^{k-1} \hat{q}_{j-1,j}\left(k-l-1\right) \hat{q}_{0,j-1} \left(l\right) &&
\end{flalign}
 for $k=0,1,...,K-1$.
\item Finally, use the estimates obtained in the last iteration of step 2 to compute
\begin{flalign}
\label{eq:est4}
\hat{p}_{0n}\left(k\right):=&\sum_{l=0}^{k-1} \hat{r}_{n}\left(k-l-1\right) \hat{q}_{0,n}\left(l\right)&&
\end{flalign}
 for $k=0,1,...,K$, where  $\hat{r}_{n}\left(\cdot\right):=\exp \left[-\hat{\Lambda}_{n}\left(t_{\scriptscriptstyle\left(\cdot\right)}\right)\right]\,$.
\end{enumerate}
Substituting $:=$ for $\approx$ and removing the `hats' in definitions \ref{eq:est1} to \ref{eq:est4},  we get the approximate equalities that justify the algorithm. These approximate equalities are derived in the Supplementary Materials (file ESM\_1.html, section 2).

Apart from \texttt{probtrans\_ebmstate}, the function \texttt{probtrans\_fft} is also based on the convolution argument just shown. 
However, this function makes use of the convolution theorem, i.e., of the fact that the convolution of two (vectorized) functions in the time domain is equivalent to a pointwise product of the same functions in the frequency domain. The estimation of state occupation probabilities is thus simplified to
\begin{align*}
 \hat{p}_{0n}:=&\mathcal{F}^{\scriptscriptstyle -1}\left\lbrace \hat{\mathrm q}_{0,1} \boldsymbol{\cdot} \hat{\mathrm q}_{1,2}\boldsymbol{\cdot} \mathrm{...}\boldsymbol{\cdot}\hat{\mathrm q}_{n-1,n}\boldsymbol \cdot \hat{\mathrm r}_{n}\right\rbrace\;, 
\end{align*}
where $\mathcal{F}$ denotes the discrete Fourier transform, $\hat{\mathrm{q}}_{j-1,j}:=\mathcal{F}(\hat{q}_{j-1,j})$ and  $\hat{\mathrm{r}}_{n}:=\mathcal{F}(\hat{r}_{n})$.
Conversion to and from the frequency domain is carried out using the fast Fourier transform algorithm implemented in the \texttt{fft} function of the \texttt{stats} package.
The Supplementary Materials contain a short simulation study checking that state occupation probabilities can be accurately estimated with \texttt{probtrans\_ebmstate} and \texttt{probtrans\_fft}  (see file ESM\_1.html, sections 3 and 4).

\begin{figure}[h] 
\centering         
\includegraphics[width=13.5cm, angle=0]{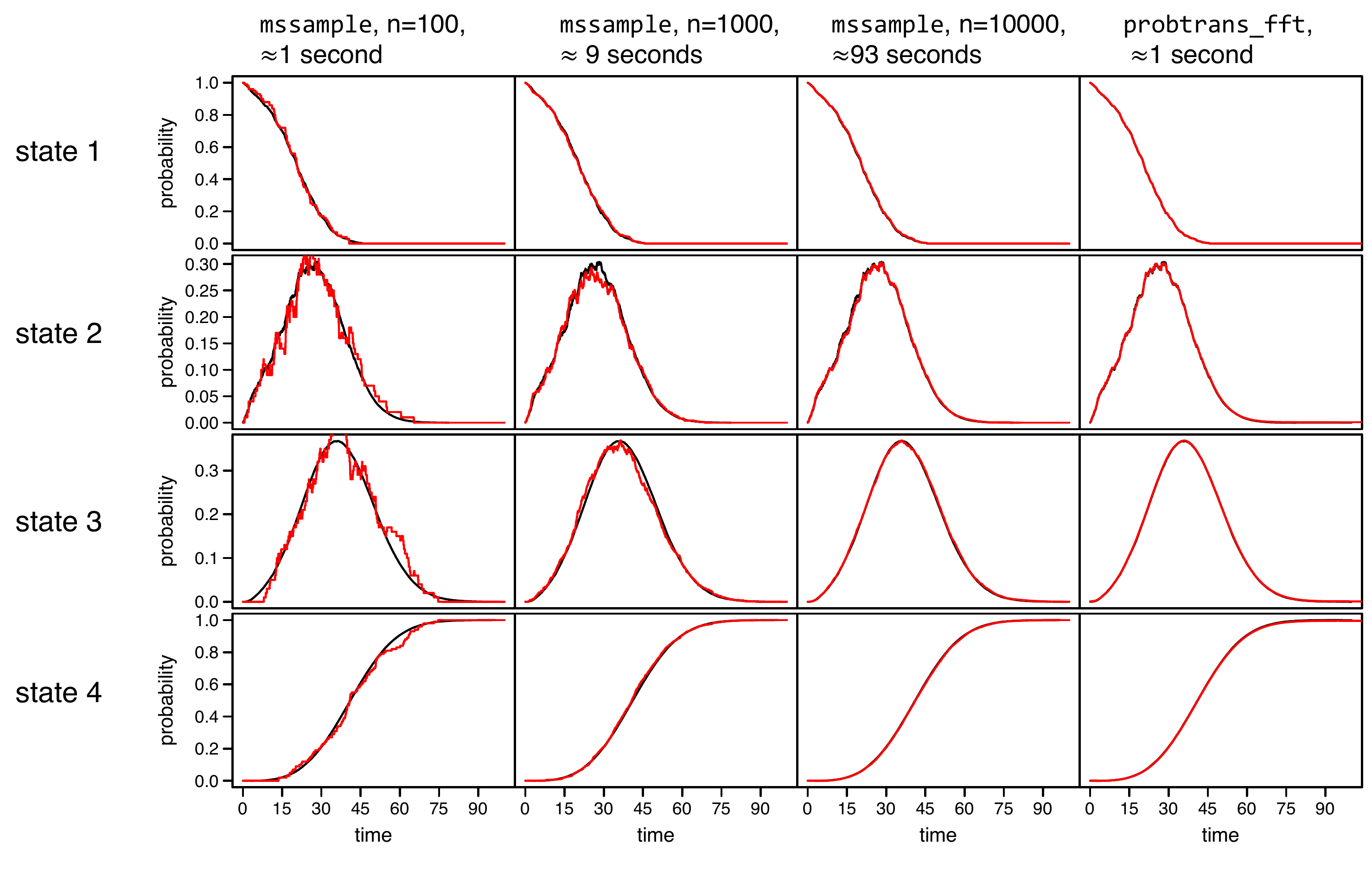} % width changes size
\vspace*{0.25cm}     % manual adjustment of vertical spacing
\caption{Comparison of running times and estimation accuracy of \texttt{mssample} and \texttt{probtrans\_fft}. Each plot in the grid shows two estimated curves of state occupation probabilities. The black curves are based on a single run of \texttt{mstate::mssample} with $n=100.000$ observations (approximately 17 minutes of running time) and are the same across columns. They serve as benchmark for precision assessment. In columns 1 to 3 of the grid, the superimposed red curves are based on a run of \texttt{mssample} with respectively 100, 1000, and 10.000 observations. In the rightmost column,  the red curves are based on a run of \texttt{probtrans\_fft}. All functions have as input the same set of cumulative transition hazards. These were estimated using a non-parametric multi-state model and a data set of 1000 patients generated according to a clock-reset Cox model with a `linear' transition structure (leftmost diagram of figure \ref{fig:transition_structures}).  Plots in the same row refer to the same state of the model, while those in the same column refer to the same run of a function. Running times and, where appropriate, number of simulations ($n$) are given on top of each column.}
\label{fig:mssample} % label for the figure
\end{figure}

Figure \ref{fig:mssample} consists of a grid of plots with estimated curves of state occupation probabilities. It compares, in terms of speed and accuracy,  the estimator in \texttt{probtrans\_fft} with an estimator in \texttt{mstate::mssample} that has the same target, but is simulation-based.  Each plot contains a black curve and a superimposed red curve.  The red curves in any given column of the grid are all based on the same run of a function: columns 1 to 3 are based on runs of \texttt{mssample} with the number of samples $n$ equal to $100$, $1000$ and $10.000$ respectively, while column 4 is based on a run of \texttt{probtrans\_fft}.  Each column in the grid reproduces the same 4 black curves. These are based on a single run of \texttt{mssample} with $n=100.000$ and serve as benchmark. All function runs are based on the same input: a set of cumulative transition hazard estimates for a multi-state model with the `linear' transition structure given in the leftmost diagram of figure \ref{fig:transition_structures}. Plots in a given row refer to the same state of the model. The running times on top of each column refer to the estimation of red curves. 
The main conclusion suggested by this analysis of simulated data is that \texttt{probtrans\_fft}  is as accurate as \texttt{mssample} with $n=10.000$, but it is almost 100 times faster (columns 3 and 4).  With $n=1000$, \texttt{mssample} achieves a good approximation to the true state occupation probabilities, but is still roughly 9 times slower.  The details on how figure \ref{fig:mssample} and its underlying data were generated are given in the Supplementary Materials (file ESM\_1.html, section 5).

\subsection{Interval estimation}
\label{sec:interval_estimation}

Under any model estimated by \texttt{ebmstate} -- as in general under a Bayesian model --,  one can, if the sample size is large enough, approximate the posterior by a normal distribution with mean equal to the maximum \textit{a posteriori} estimate and covariance matrix equal to the inverse of the generalised observed Fisher information \citep[see, for example,][p. 83-84]{Gelman2014}. This approximation has first-order accuracy and is thus outperformed by Laplace's method, which has second-order accuracy \citep[][p. 110-111]{Carlin2009}. However, as \citet[p. 112]{Carlin2009} observe, ``for moderate- to high-dimensional $\boldsymbol\theta$ (say, bigger than 10), Laplace\textquotesingle s method will rarely be of sufficient accuracy[...]''. 
\citet[][p. 244-251]{Carlin2009} also describe three methods of interval estimation in empirical Bayes settings, but all of them are designed for fully parametric models. These reasons, along with the fact that regularised methods such as the one implemented  \texttt{ebmstate} are typically  used to fit models with more than a dozen covariates, led us to choose the non-parametric bootstrap as the interval estimation method in \texttt{ebmstate}. Interval estimates of regression coefficients, cumulative hazards and state occupation probabilities are implemented in the function \texttt{boot\_ebmstate}.

\section{Estimator performance}
\label{sec:estimator_performance}
It is a well-documented fact in the statistical literature that standard least-squares or maximum-likelihood estimators can often be improved by regularisation or shrinkage \citep[see, for example,][]{Samworth2012}. This improvement comes about when the model dimensionality is high enough that the bias introduced by regularisation is outweighed by the reduction in the estimator variance.  In the current setting, one might therefore ask: what kind of dimensionality does a semi-parametric, multi-state Cox model need to have to be outperformed by its empirical Bayes counterpart? 
 A simulation study we carried out offers a tentative answer to this question, by comparing estimators under both Cox models for an increasing number of covariates.  The study also features a third method, based on a fully non-parametric model, as a null model method.  This was included to give an idea of how many covariates the empirical Bayes 
model can deal with before it becomes no better than a simple non-regressive model.

\subsection{Simulation setup}
\begin{figure}[h] 
\centering         
\includegraphics[width=14.5cm, angle=0]{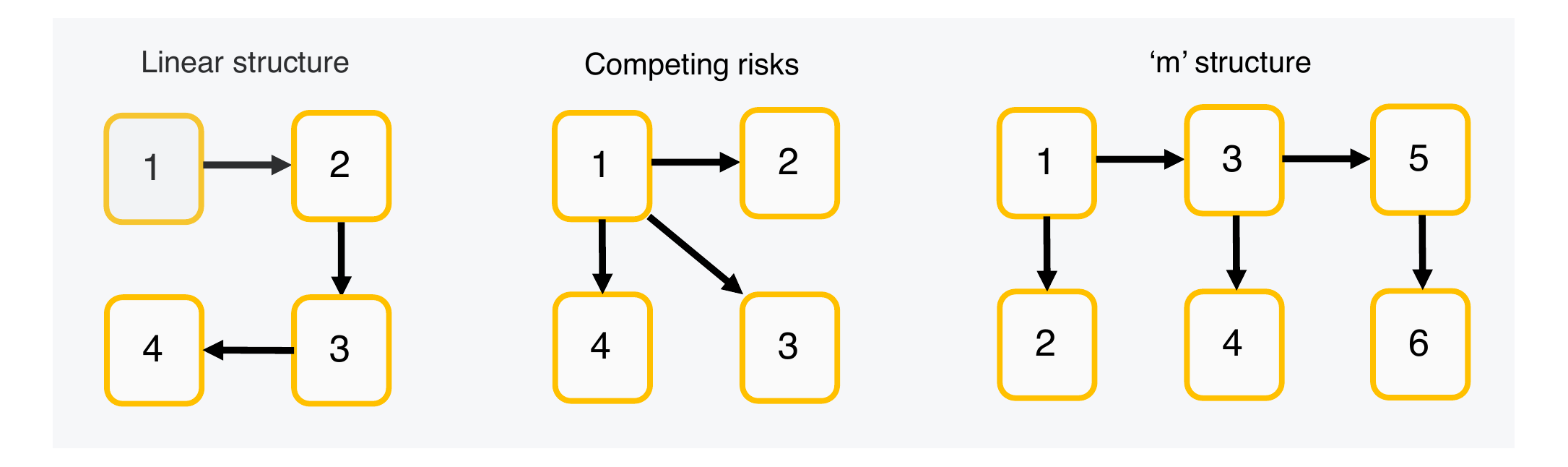} % width changes size
\vspace*{0.25cm}     % manual adjustment of vertical spacing
\caption{Model transition structures. We studied the performance of Cox model estimators, empirical Bayes Cox  model estimators and fully non-parametric estimators with respect to these 3 transition structures.}      
\label{fig:transition_structures} % label for the figure
\end{figure}

We assessed the performance of all estimators defined by the tuple $\left[a,m, G, n,p(n)\right]$, where $a\in \lbrace$regression coefficients, relative hazards, state occupation probabilities$\rbrace$ is the target of estimation,  $m\in \lbrace$standard Cox, empirical Bayes Cox, null$\rbrace$ is the assumed hazard model,  $G \in \lbrace$linear, competing risks, `m' structure$\rbrace$ is the transition structure of the model (illustrated in figure \ref{fig:transition_structures}) and $n\in \lbrace 100,1000\rbrace$ is the number of patients/disease histories in the training data set;
the variable $p$ denotes the number of coefficients/covariates per transition in the true model and its range depends on $n$: $p\left(100\right) \in \lbrace 10,40,70,100 \rbrace$ whereas $p\left(100\right) \in \lbrace 10,100,200,300 ,400,500\rbrace$.  By `relative hazards' and `state occupation probabilities', we mean here the relative transition hazards of an out-of-sample patient, and her state occupation probabilities at 7 chosen time points.
We generated a batch of 300 independent absolute error observations (`NA' estimates included) for each estimator, where each observation is recorded after training the estimator on a newly simulated data set.  Each boxplot in figures \ref{fig:estimator_performance_boxplots_100patients} ($n=100$) and \ref{fig:estimator_performance_boxplots_1000patients} ($n=1000$) is based on one of these batches. As all estimators are \textit{vector} estimators,  each absolute error is actually an \textit{average} absolute error, where the average is taken over the components of the vector.

All training data sets were simulated from clock-reset Cox models. Apart from $G$ (the model transition structure), $n$ and $p$, also the true baseline hazards are held fixed within each batch of 300 training data sets.  
%However, the regression coefficient vector of the model is sampled at random before simulating a new data set, as we found that using the same vector in the simulation of all data sets in a batch would make the results highly dependent on the actual coefficient values used.
The coefficient vectors used in the simulation are always non-sparse and are scaled by $\sqrt{\frac{10}{p}}$ to keep the log-hazard variance constant when the dimensionality grows.  
All covariates are dichotomous and mutually independent.
To compute the coefficient errors for the non-parametric (null) model method, we think of it as a degenerate Cox model in which all regression coefficient estimates are fixed at zero. 
The estimation of regression coefficients under the standard Cox and the empirical Bayes Cox models was performed with \texttt{survival::coxph} and \texttt{ebmstate::CoxRFX} respectively; the estimation of state occupation probabilities is based on \texttt{mstate::probtrans} for the null model and on \texttt{ebmstate::probtrans\_fft} for both the standard Cox and the empirical Bayes Cox models.

 The reason we did not consider simulation scenarios with more than 500 covariates per transition, in data sets of 1000 patients, was simply computational cost. For example, generating the data and error observations for the scenario with $n=1000$, $p=100$ and $G=$`m' structure took less than one hour to generate using 20 CPU cores in parallel; the same scenario but with $p=500$ took 6.5 days using 25 CPU cores. More details about the simulation setup can be found in the Supplementary Materials (file ESM\_1.html, section 6,  subsection `sample script').

\subsection{Missing values}
\begin{figure}[h] 
\centering         
\includegraphics[width=14.5cm, angle=0]{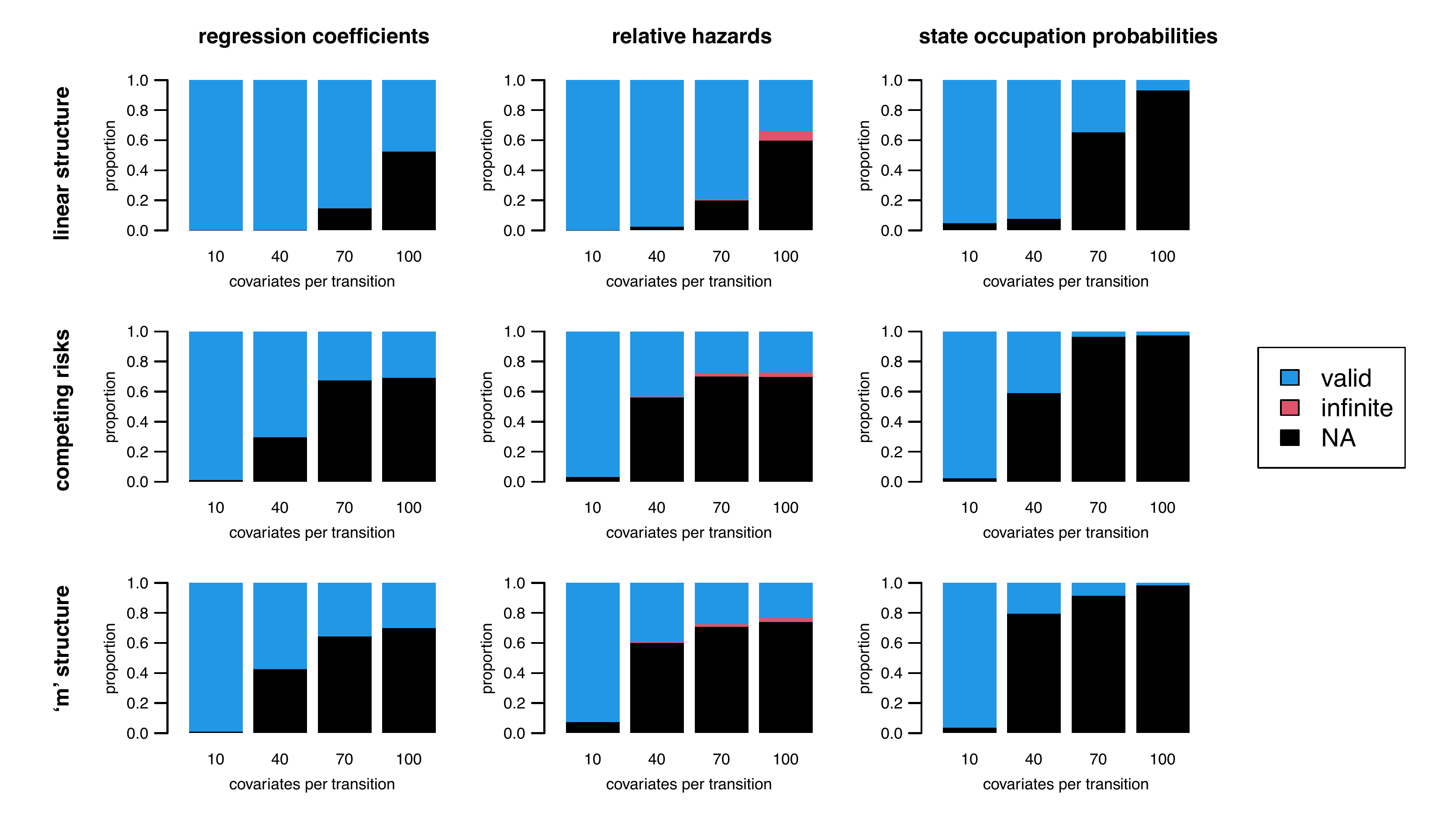} % width changes size
\vspace*{0.25cm}     % manual adjustment of vertical spacing
\caption{Proportions of valid, infinite and missing (`NA') estimates for the standard Cox model estimators in the simulation study of figure \ref{fig:estimator_performance_boxplots_100patients} (100 patients per simulated data set).}  
\label{fig:na_props_100patients_coxph} % label for the figure
\end{figure}

\begin{figure}[h] 
\centering         
\includegraphics[width=14.5cm, angle=0]{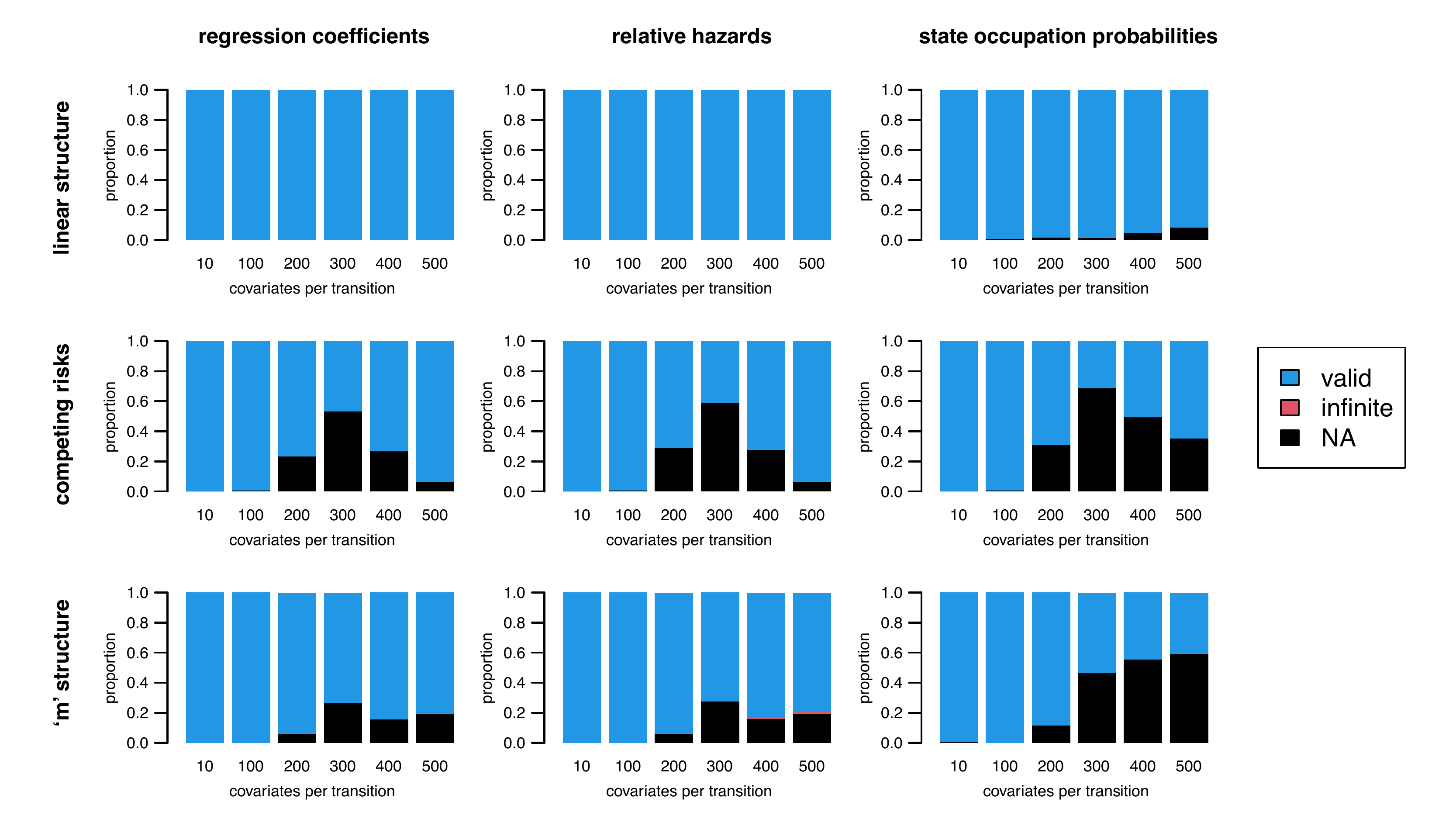} % width changes size
\vspace*{0.25cm}     % manual adjustment of vertical spacing
\caption{Proportions of valid, infinite and missing (`NA') estimates for the standard Cox model estimators in the simulation study of figure \ref{fig:estimator_performance_boxplots_1000patients} (1000 patients per simulated data set).}
\label{fig:na_props_1000patients_coxph} % label for the figure
\end{figure}

Whenever an estimator was able to compute a valid estimate of its target for each training data set,   i.e., when it did not return any `NA' estimates,  its boxplots are based on 300 valid error observations.  This was always the case with non-parametric estimators: the estimates of regression coefficients and relative hazards of this type of estimators are trivial (fixed at zero and one respectively) and hence it is also straightforward to compute absolute errors. 
 It also happened that non-parametric estimators of state occupation probabilities had no `NA' estimates (see file ESM\_1.html, section 6, figure 6.3, in the Supplementary Materials). The situation was similar for the empirical Bayes Cox model estimators, which showed no more than 5$\%$ missing estimates in any of the simulation scenarios  studied (ibid., figures 6.1 and 6.2).  However, for the standard Cox model ones, the number of `NA' estimates  depends to a large extent on the number of patients in the data set, as well as on the dimensionality and transition structure of the model (figures \ref{fig:na_props_100patients_coxph} and \ref{fig:na_props_1000patients_coxph}). In data sets of 100 patients, it fares well in models with fewer than 10 covariates per transition, or in models with up to 40 covariates, if the transition structure is linear. Otherwise its failure rates range from roughly 25$\%$ to nearly 100$\%$. 
In data sets of 1000 patients, the proportion of `NA' estimates is never above 10$\%$, if the transition structure is linear, but it can climb above 60$\%$ for other transition structures.

\subsection{Comparison of estimators}
\begin{figure}[h] 
\centering         
\includegraphics[width=14.5cm, angle=0]{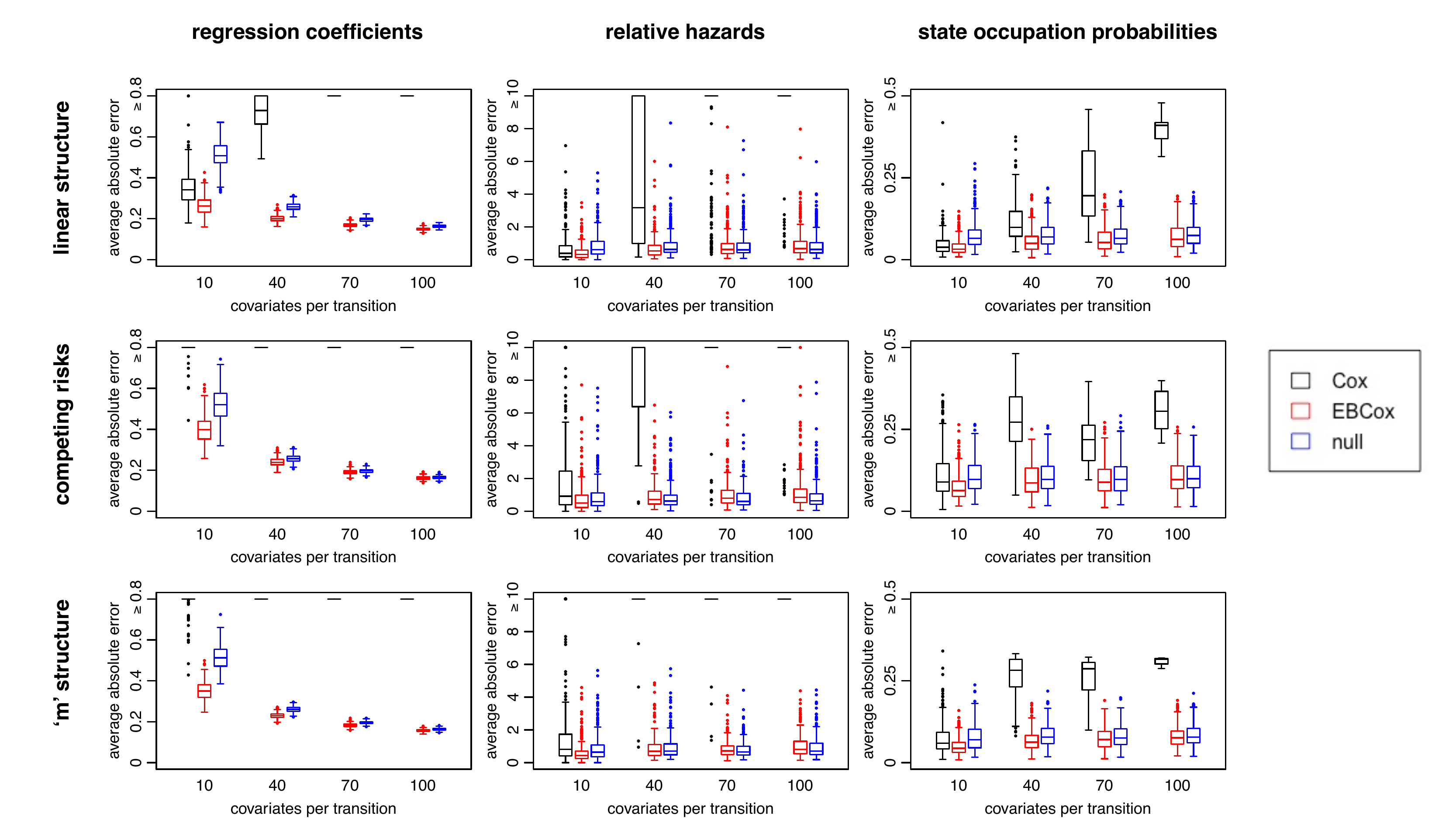} % width changes size
\vspace*{0.25cm}     % manual adjustment of vertical spacing
\caption{Performance comparison of standard Cox,  empirical Bayes Cox,  and fully non-parametric (null) estimators using training data sets with \textbf{100 observations} each.  In the figure grid there is a boxplot corresponding to every tuple $(a,m, G, p)$ such that $a\in \lbrace$regression coefficients, relative hazards, state occupation probabilities$\rbrace$ is the target of estimation,  $m\in \lbrace$standard Cox, empirical Bayes Cox, null$\rbrace$ is the hazard model,  $G \in \lbrace$linear, competing risks, `m' structure$\rbrace$ is the transition structure of the model, and $p \in \lbrace 10,40,70,100 \rbrace$ is the number of coefficients/covariates per transition.  
Each boxplot is based on at most 300 average absolute error observations.
Figure \ref{fig:na_props_100patients_coxph}, together with figures 6.1 and 6.3 in file ESM\_1.html of the Supplementary Materials, show the proportion of valid, missing and infinite estimates for each estimator. In each simulation scenario, the upper limit of the plot's y-axis  defines a threshold above which observations are considered very large. Very large observations were replaced by the y-axis upper limit before the boxplots were built. 
}      
\label{fig:estimator_performance_boxplots_100patients} % label for the figure
\end{figure}

\begin{figure}[h] 
\centering         
\includegraphics[width=14.5cm, angle=0]{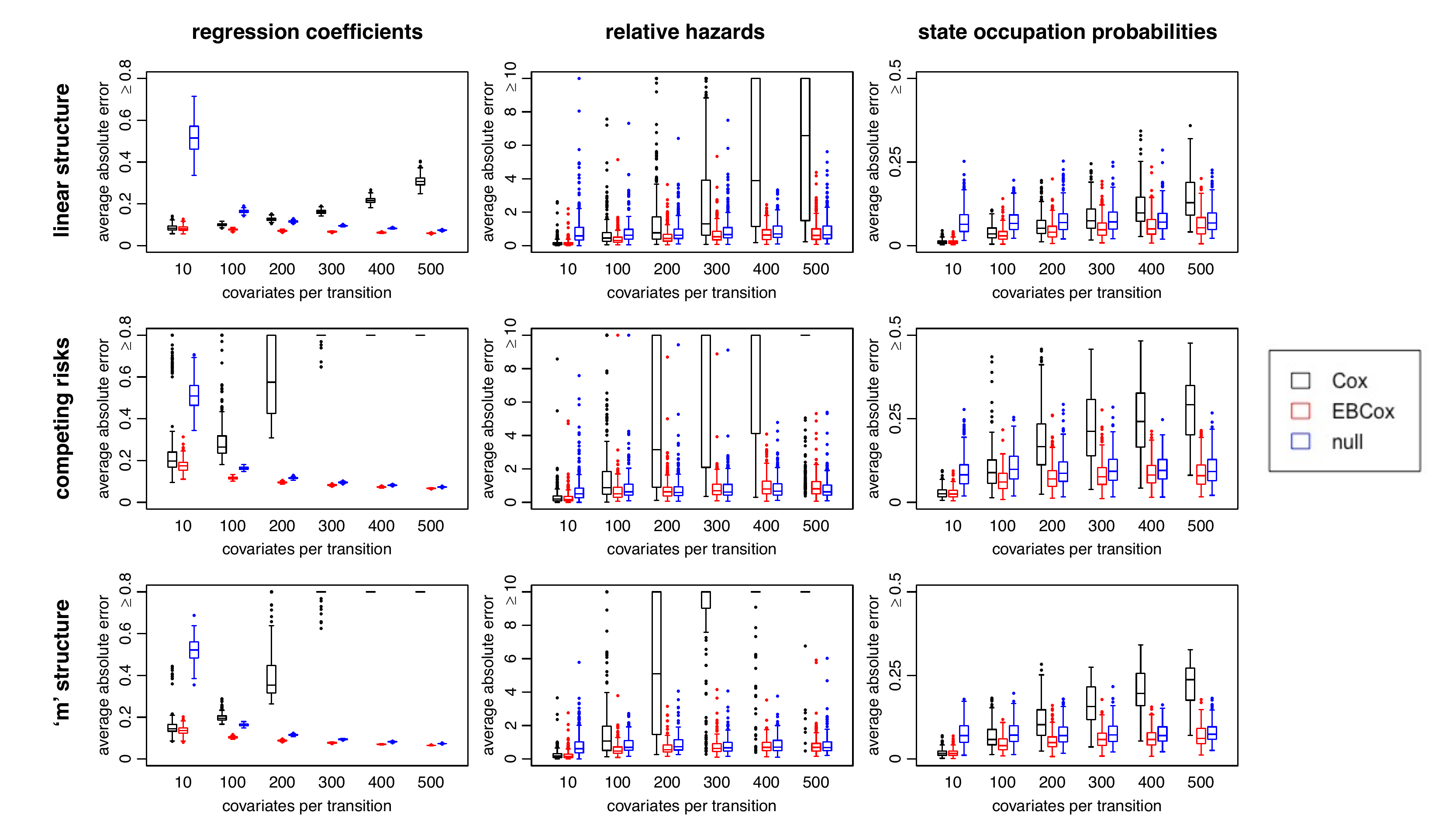} % width changes size
\vspace*{0.25cm}     % manual adjustment of vertical spacing
\caption{
Performance comparison of standard Cox,  empirical Bayes Cox,  and fully non-parametric (null) estimators using training data sets with \textbf{1000 observations} each.  In the figure grid there is a boxplot corresponding to every tuple $(a,m, G, p)$ such that $a\in \lbrace$regression coefficients, relative hazards, state occupation probabilities$\rbrace$ is the target of estimation,  $m\in \lbrace$standard Cox, empirical Bayes Cox, null$\rbrace$ is the hazard model,  $G \in \lbrace$linear, competing risks, `m' structure$\rbrace$ is the transition structure of the model, and $p \in \lbrace 10,100,200,300,400,500 \rbrace$ is the number of coefficients/covariates per transition.  
Each boxplot is based on at most 300 average absolute error observations. 
Figure \ref{fig:na_props_1000patients_coxph}, together with figures 6.2 and 6.3 in file ESM\_1.html of the Supplementary Materials, show the proportion of valid, missing and infinite estimates for each estimator. In each simulation scenario, the upper limit of the plot's y-axis  defines a threshold above which observations are considered very large. Very large observations were replaced by the y-axis upper limit before the boxplots were built. 
}      
\label{fig:estimator_performance_boxplots_1000patients} % label for the figure
\end{figure}

With respect to the performance of the three methods studied, the boxplots in figures \ref{fig:estimator_performance_boxplots_100patients} and \ref{fig:estimator_performance_boxplots_1000patients} suggest the following conclusions:
\begin{itemize}
\item As $p/n$ grows,  the empirical Bayes estimators quickly outperform the standard Cox model ones.  They already fare substantially better at $p/n=0.1$ for both $n=100$ and $n=1000$ and for all estimation targets.  At the same time, the relative performance of the empirical Bayes method with respect to the null model one decreases.  At $p/n=0.5$, the difference between these two methods is already rather small for all simulation scenarios. 
\item The relative performance of the empirical Bayes method with respect to the null method decreases as the number of co-occurring transition hazards in the model grows.  All other things equal, the empirical Bayes method has the best performance under the `linear' structure model, which has no competing transitions; it performs less well under the `m' structure transition model, where two transition hazards can co-occur; and has the worse relative performances under the `competing risks' model, where three transition hazards co-occur.  This trend is clearer for $n=100$ (figure \ref{fig:estimator_performance_boxplots_100patients}) but can also be detected in the relative hazard errors for $n=1000$ (figure \ref{fig:estimator_performance_boxplots_1000patients}).   In any case, the empirical Bayes method seems to be far more robust than the standard Cox model against increases in the number of co-occurring transition hazards.
\item Having as target the regression coefficients or the state occupation probabilities,  instead of relative hazards, makes the empirical Bayes method better in comparison to the null method. In fact, as $p/n$ grows, the empirical Bayes method is never outperformed by the null method except in the estimation of relative hazards.
\end{itemize}

\section{SURVIVAL ANALYSIS WORKFLOW}
\begin{figure}[h] 
\centering         
\includegraphics[width=13.5cm, angle=0]{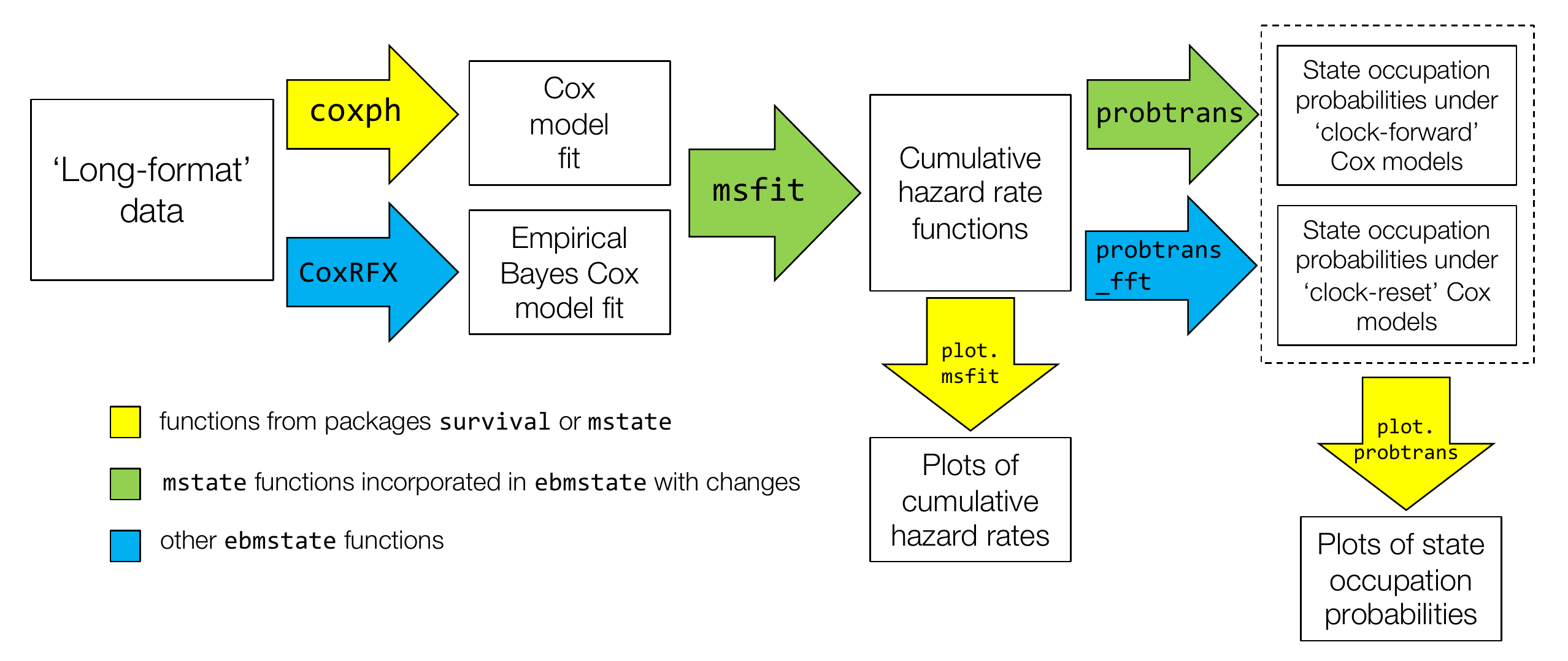} % width changes size
\vspace*{0.25cm}     % manual adjustment of vertical spacing
\caption{Extension of the \texttt{mstate} analysis framework by \texttt{ebmstate}. Arrows correspond to functions. Boxes correspond to inputs or outputs of functions. Functions \texttt{CoxRFX} and \texttt{probtrans\_fft} from \texttt{ebmstate} compute point estimates only.  Interval estimates can be obtained using the non-parametric bootstrap algorithm implemented in the function \texttt{ebmstate::boot\_ebmstate}.}      
\label{fig:workflow} % label for the figure
\end{figure}

The features of \lstinline!mstate! were illustrated in  \citet{Wreede2010} using a simple workflow. The starting point of this workflow is a data set in `long format'. Such data set can be fed into \lstinline!survival::coxph! to obtain estimates of the regression coefficients of a multi-state Cox model. The resulting model fit object can be passed on to  \lstinline!mstate::msfit!, along with a vector of covariates of a particular patient, to get personalised estimates of the cumulative hazard functions. 
Finally, state occupation probabilities for the same patient can be estimated if the object created by \lstinline!mstate::msfit! is fed into  \lstinline!mstate::probtrans!. 
In this section, we describe how \texttt{ebmstate} extends the scope of this workflow, i.e., how it uses the packages \texttt{survival} and \texttt{mstate} to generate estimates under a multi-state empirical Bayes Cox model.  A diagram summarising the extension is shown in figure \ref{fig:workflow}. 

\begin{figure}[h] 
\centering         
\includegraphics[width=14.5cm, angle=0]{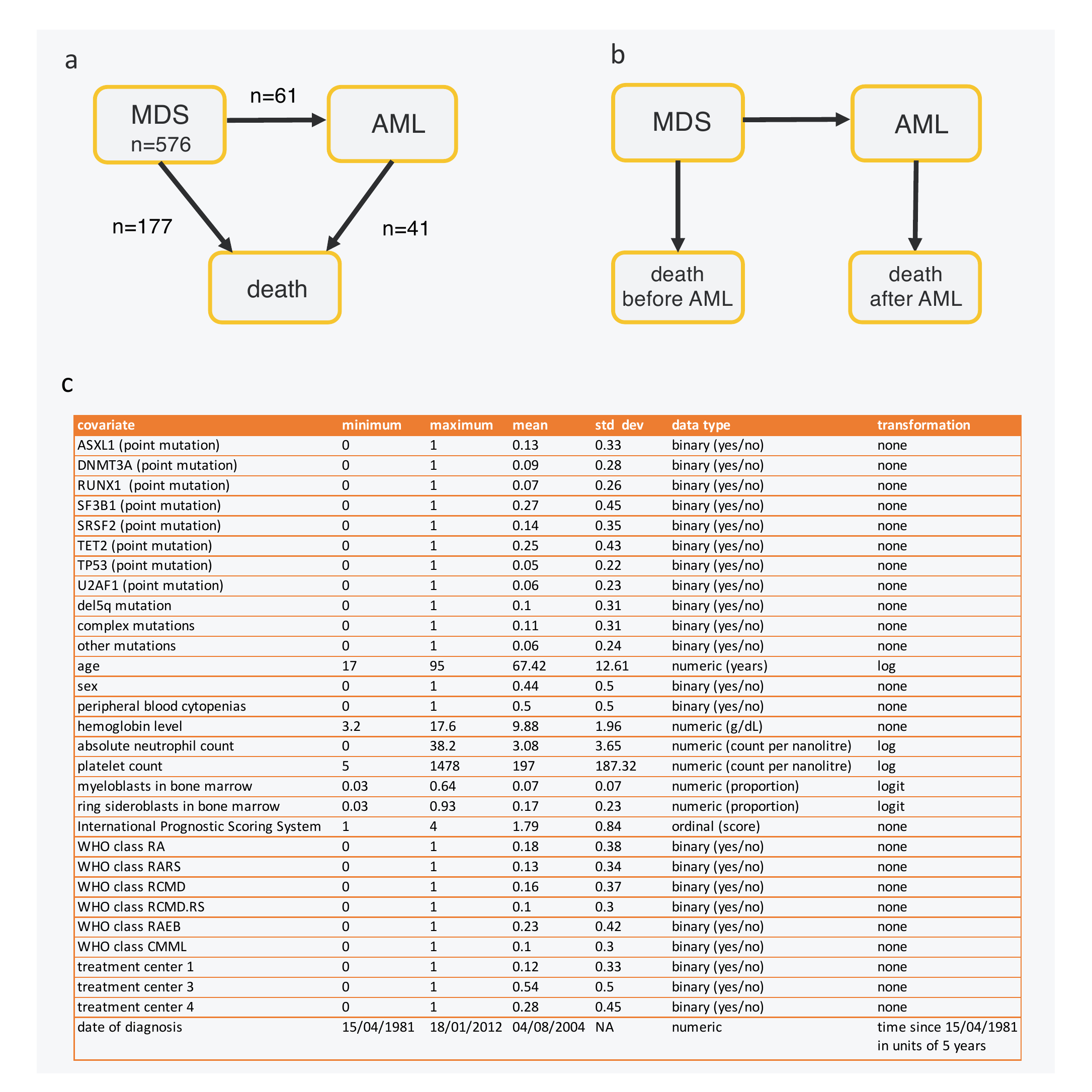} % width changes size
\vspace*{-0.25cm}     % manual adjustment of vertical spacing
\caption{\textbf{a}: transition model implied by the data set of patients with myelodysplastic syndromes, together with transition event numbers; \textbf{b}: conversion to a tree-like transition structure; \textbf{c}: transformations applied to the MDS covariate data and summary statistics for the data before transformation. MDS stands for \textit{myelodysplastic syndromes}; AML stands for \textit{acute myeloid leukemia}.}      
\label{fig:trans_diagrams} % label for the figure
\end{figure}

 The main steps of the \texttt{ebmstate} workflow will be illustrated using a data set of patients with myelodysplastic syndromes (MDS) which has been described and studied in \citet{Papaemmanuil2013}. A myelodysplastic syndrome is a form of leukemia in which the bone marrow is not able to produce enough mature blood cells, and which sometimes develops into a cancer of white blood cells with a quick and aggressive progression, i.e., into acute myeloid leukemia (AML). Figure \ref{fig:trans_diagrams}a illustrates an illness-death type model for MDS patients and also gives a breakdown of the number of transition events. 
 The conversion to a model with a tree-like transition structure (that can be handled by our convolution-based estimators) is shown in figure \ref{fig:trans_diagrams}b. The data set used for model estimation, obtained after a number of pre-processing steps,  contains the disease history of 576 patients, as well as measurements on 30 covariates. Of these 30 covariates, 11 are mutation covariates and the remaining are clinical or demographic (see figure \ref{fig:trans_diagrams}c). 
 The running time for the estimation of relative transition hazards does not exceed 10 seconds in a standard laptop computer. The same holds for the estimation of  cumulative transition hazards or state occupation probabilities for a given patient. The complete R code underlying the data analysis in the current section can be found in the Supplementary Materials (file ESM\_2.html).

\subsection{Input data}
Table \ref{table:long_format_data} shows a fragment of the MDS data set. The data is in `long format', which means that each row refers to a period of risk for a given transition and patient. 
For example, row $i$ tells us that, at time \texttt{Tstart[i]}, patient \texttt{id[i]} entered state \texttt{from[i]}, and thereby began to be at risk  for transition \texttt{trans[i]}, i.e., at risk of going from state \texttt{from[i]} to state \texttt{to[i]}. If the first transition of  patient \texttt{id[i]} after time \texttt{Tstart[i]}   occurs  before the last follow-up time for this patient, \texttt{Tstop[i]} records the time of this transition (regardless of whether the patient moved to state \texttt{to[i]} or not). Otherwise, \texttt{Tstop[i]} is set to the last follow-up time. The value of  \texttt{status[i]} is set to 1 if and only if the first transition of patient \texttt{id[i]}  after \texttt{Tstart[i]} is to state \texttt{to[i]} and occurs before the last follow-up (otherwise it is set to 0). 
The value of  \texttt{time[i]} is defined simply as $\texttt{Tstop[i]} - \texttt{Tstart[i]}$, and \texttt{strata[i]} is the stratum of the baseline hazard for transition \texttt{trans[i]} (more about this variable in the following section). For $\texttt{x} \in \left\lbrace \texttt{ASXL1}, \texttt{DNMT3A}, \dots \right \rbrace$,  \texttt{x[i]} denotes the level of covariate \texttt{x} between \texttt{Tstart[i]} and \texttt{Tstop[i]} in patient \texttt{id[i]}. (In the MDS data set, we assume that the relative hazard of a patient is determined by her covariate vector at $t=0$, i.e., we assume all covariates to be time-fixed.) 
If a patient enters a new state, and this state communicates directly with $n$ other states, then, as long as the patient actually spends time in the new state (i.e. the time of transition is not the same as the last follow-up time), $n$ rows must be added to the data set, with each row corresponding to a different possible transition.

From table \ref{table:long_format_data}, we know that patient 1 entered state 1 (`MDS') at time 0 and remained in this state until time 327, when she moved to state 3 (`death before AML'). There are no rows to describe the evolution of patient 1 after entering state 3, as this state is an absorbing state. As to patient 2, she remained in state 1 until time 1613, and moved from there to state 2 (`AML'). By the time of the last follow-up (1782), patient 2 was still in state 2 (and still at risk of going to state 4 -- `death after AML').

\begin{table}
%The file whose content is printed below was generated in R with code
% like the following:
%> sink("long_data.txt")
%> print(long_data[1:5,1:5])
%> sink()
\small
\begin{lstlisting}
  id from to trans Tstart Tstop time status  strata ASXL1 DNMT3A [...]  
1  1    1  2     1      0   327  327      0       1     0      0    .
2  1    1  3     2      0   327  327      1       2     0      0    .
3  2    1  2     1      0  1613 1613      1       1     1      0    .
4  2    1  3     2      0  1613 1613      0       2     1      0    .
5  2    2  4     3   1613  1782  169      0       3     1      0    .
\end{lstlisting}
\caption{The first five rows of the MDS data set (in long format)}
\label{table:long_format_data}
\end{table}

\subsection{Fitting an empirical Bayes Cox model}
\label{sec:fit_bayes_cox_model}
Once the data is in `long format', the estimation of an empirical Bayes model can be carried out using the function \texttt{CoxRFX}. A simple example of the first argument of \texttt{CoxRFX}, denoted `\texttt{Z}', is a data frame gathering the \texttt{trans}, \texttt{strata} and covariate columns of the data in long format:
\small
\begin{lstlisting}
Z<-mstate.data[!names(mstate.data)%in%c("id","from","to",
	"Tstart","Tstop","time","status")]  
#(`mstate.data' has the data in long format) 
\end{lstlisting}
\normalsize
The \texttt{strata} column  determines which baseline hazard functions are assumed to be equal.
 In table \ref{table:long_format_data}, each transition is assumed to have a (potentially) different baseline hazard. The model's assumptions regarding how covariates affect the hazard are reflected on the format of the covariate columns of \texttt{Z}. When the \texttt{Z} argument is the one created in the previous block of code, \texttt{CoxRFX} returns a single regression coefficient estimate for each covariate. In other words, the impact of any covariate  is assumed to be the same for every transition.

There is however a way of relaxing this assumption. One can replace the \texttt{ASXL1} column in Z (or any other covariate column) by several `type-specific' \texttt{ASXL1} columns: the \texttt{ASXL1} column specific for  type $i$  would show the mutation status of \texttt{ASXL1}  in rows belonging to transition of type $i$, and show zero in all other rows. This would force \texttt{CoxRFX} to estimate a (potentially) different \texttt{ASXL1} coefficient for each  transition type.  This process of covariate expansion by type can be based on any partition of the set of transitions. When each type corresponds to a single transition,  we refer to it simply as `covariate expansion by transition'. The output shown below illustrates the effect of expanding the covariates in `mstate.data' by transition.
\begin{minipage}{\linewidth}
\small
\begin{lstlisting}
# Columns `id' and `trans' from `mstate.data' together with the first
# two expanded covariates (first 5 rows only):
    id trans ASXL1.1 ASXL1.2 ASXL1.3 DNMT3A.1 DNMT3A.2 DNMT3A.3  [...]  
1    1     1       0       0       0        0        0        0     .
2    1     2       0       0       0        0        0        0     .
3    2     1       1       0       0        0        0        0     .
4    2     2       0       1       0        0        0        0     .
5    2     3       0       0       1        0        0        0     .
\end{lstlisting}
\normalsize
\end{minipage}
The example code given below shows how to use \texttt{mstate} to expand covariates by transition and how to create a \texttt{Z} argument that makes \texttt{CoxRFX} estimate a regression coefficient for each covariate for transitions 1 and 2, and assume a fully non-parametric hazard for transition 3.
\small
\begin{lstlisting}
# To expand covariates by transition using mstate::expand.covs, 
# first set the class of `mstate.data' as
class(mstate.data)<-c("data.frame","msdata")

# then add the transition matrix as attribute:
attr(mstate.data,"trans")<-tmat #`tmat' is the output of mstate::transMat

# Expand covariates by transition:
covariates.expanded<-mstate::expand.covs(mstate.data,covs = 
	names(mstate.data)[!names(mstate.data)%in%c("id",
	"from","to","trans","Tstart","Tstop","time","status","strata")],
	append = F)

# remove all covariates for transition 3 from `covariates.expanded'
# to fit a fully non-parametric model on this transition:
covariates.expanded<-covariates.expanded[!grepl(".3", 
	names(covariates.expanded),fixed = T)]

#argument `Z' of coxrfx
Z<-data.frame(covariates.expanded,strata=mstate.data$trans, 
	trans=mstate.data$trans)
\end{lstlisting}
\normalsize

The second argument of \texttt{CoxRFX} (`\texttt{surv}') is a survival object that can easily be built by feeding the outcome variable columns of the data to the function \texttt{Surv} (from the  package \texttt{survival}). 
Whether \texttt{CoxRFX} fits a clock-forward model or a clock-reset model depends on the kind of survival object:
\small
\begin{lstlisting}
#argument `surv' for a  clock-forward model
surv<-Surv(mstate.data$Tstart,mstate.data$Tstop,mstate.data$status)

#argument `surv' for a clock-reset model
surv<-Surv(mstate.data$time,mstate.data$status)
\end{lstlisting}
\normalsize

The argument \texttt{groups} of \texttt{CoxRFX} is a vector  whose length equals the number of covariates in the data. In other words, the length of \texttt{groups} is \texttt{ncol(Z)-2}, since the argument \texttt{Z} must include both the covariate data and the \texttt{strata} and \texttt{trans} columns. If, for $i \neq j $, \texttt{groups[i]}=\texttt{groups[j]} $=\text{`foo'}$, this means that the regression coefficients of the $i^{th}$ and $j^{th}$ covariates of \texttt{Z} both belong to a group named `foo' of coefficients with the same prior. For the \texttt{Z} object built above, the \texttt{groups}  argument created in the following block of code embodies the assumption that all coefficients associated with a given transition have the same prior distribution. The final line of code fits the empirical Bayes model.
\small
\begin{lstlisting}
#argument `groups' of coxrfx
groups<-paste0(rep("group",ncol(Z)-2),c("_1","_2"))

#fit random effects model
coxrfx_object<-CoxRFX(Z,surv,groups,tmat)
\end{lstlisting}
\normalsize

\begin{figure}[h] 
\centering         
\includegraphics[width=12.5cm, angle=0]{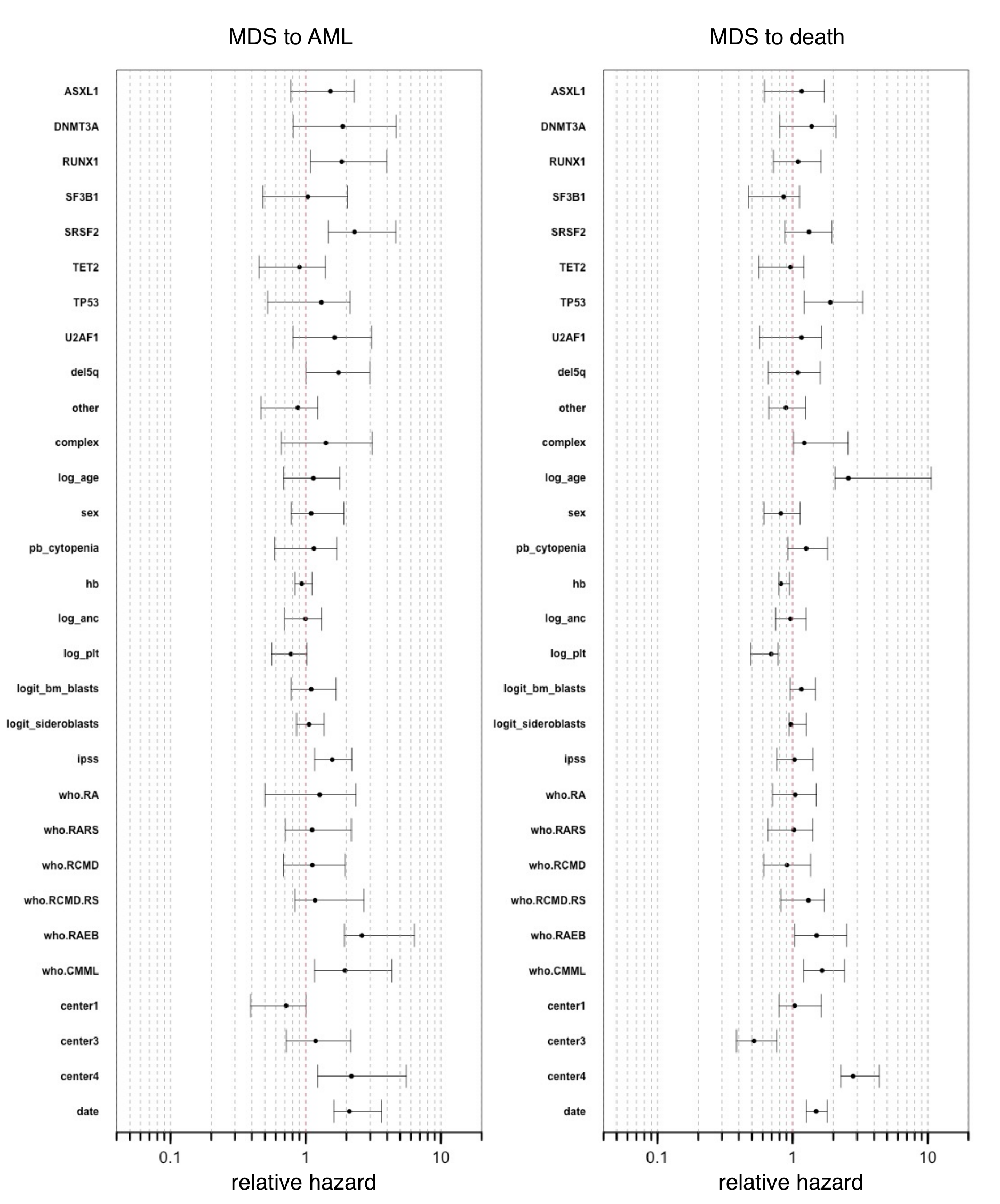} % width changes size
\vspace*{-0.25cm}     % manual adjustment of vertical spacing
\caption{Point estimates of regression coefficients for the Cox model fitted to the MDS data, along with 95\% non-parametric bootstrap confidence intervals. The $x$-axis scale is logarithmic so that coefficient estimates can be read as relative hazard estimates.  If $\gamma_{ij}$ is the element of $\hat{\boldsymbol{\beta}}_{ij}$ associated with a given covariate, $\exp\left(\gamma_{ij}\right)$ is the estimated relative hazard for this covariate in transition $\left(i,j\right)$. In general, a relative hazard estimate $r$ for a covariate $z$ in transition $\left(i,j\right)$ means that a one-unit increase in $z$ is associated with an $r$-fold increase in the hazard of this transition. If $z$ was obtained by log-transformation (as in age, platelet counts and neutrophil counts), a one-unit increase in $z$ corresponds to scaling the original covariate by $e\approx 2.72$. In case $z$ was obtained by logit-transformation (as in bone marrow blasts and sideroblasts proportions), the same one-unit increase corresponds to scaling the odds of the original covariate by $e$.}      
\label{fig:coef_plots} % label for the figure
\end{figure}

Figure \ref{fig:coef_plots} shows regression coefficient point estimates for a clock-reset, empirical Bayes model fitted with the code above. Also shown are 95\% non-parametric bootstrap confidence intervals computed using \texttt{ebmstate::boot\_ebmstate}. The $x$-axis scale is logarithmic to allow estimates to be read as relative hazards more easily.  For example, a mutation in \textit{RUNX1} is associated with a twofold increase in the hazard of progression from MDS to AML, and treatment centre 4 is associated with a 3-fold increase in the hazard of dying before progressing to AML, when compared to the baseline value of `treatment centre' (treatment centre = 2 or 5). In covariates that have been log-transformed (age, platelet count and neutrophil count) or logit-transformed (proportions of myeloblasts and ring sideroblasts in the bone marrow), the interpretation of estimates is different. For example, an increase in age by a factor of $e$ ($\approx 2.72$) almost triples the hazard of dying before AML; the same increase in the ratio $bm\_blasts/(1-bm\_blasts)$ (where \textit{bm\_blasts} is the  proportion of myeloblasts in the bone marrow) is associated with an increment in the hazard of dying before AML of approximately $16\%$.

\subsection{Computing cumulative transition hazard estimates}
\label{sec:computing_cumulative_hazards}
The function \texttt{msfit\_generic} is the generic function in \texttt{ebmstate} that computes cumulative transition hazards for a given set of covariate values and an estimated Cox model.  It calls a different method according to the class of its \texttt{object} argument. The default method corresponds to the original \texttt{msfit} function of the \texttt{mstate} package and is appropriate for objects of class \texttt{coxph}, i.e., objects that contain the fit of a Cox model with fixed effects. The other available method for \texttt{msfit\_generic}, \texttt{msfit\_generic.coxrfx}, is just the original \texttt{msfit} function, (slightly) adapted to deal with objects generated by \texttt{CoxRFX}. 
Quite importantly,  \texttt{msfit\_generic.coxrfx} does not allow the variance of the cumulative hazards to be computed, as this computation relies on asymptotic results which may not be valid for an empirical Bayes model. As a result, it only has two other arguments apart from the object of class \texttt{coxrfx}: a data frame with the covariate values of the patient whose cumulative hazards we want to compute; and a transition matrix describing the states and transitions in the model (such as the one that can be generated using \texttt{transMat} from the package \texttt{mstate}). 
The following block of code exemplifies how these objects can be built and generates the \texttt{msfit} object containing the cumulative transition hazard estimates for a sample patient. Note that the object with the patient data must include a row for each transition, as well as a column specifying the transition stratum of each row of covariates.
\small
\begin{lstlisting}
# Build `patient_data' data frame with the covariate values for which 
# cumulative hazards are to be computed (covariate values of patient 3):
patient_data<-mstate.data[mstate.data$id==3,,drop=F][rep(1,3),]
patient_data$strata<-patient_data$trans<-1:3
patient_data<-mstate::expand.covs(patient_data,covs = 
	names(patient_data)[!names(patient_data)%in%c("id",
	"from","to","trans","Tstart","Tstop","time","status","strata")],
	append = T)
patient_data<-patient_data[!grepl(".3",names(patient_data),fixed = T)]

# The `patient_data' data frame has only 3 rows (one for each transition).
# The output below shows its `id' and `trans' columns
# and expanded covariates ASXL1 and DNMT3A:
    id trans ASXL1.1 ASXL1.2 DNMT3A.1 DNMT3A.2  [...]  
1    3     1       1       0        0        0     .
2    3     2       0       1        0        0     .
3    3     3       0       0        0        0     .

# compute cumulative hazards
msfit_object<-msfit_generic(coxrfx_object,patient_data,tmat)
\end{lstlisting}
\normalsize

\begin{figure}[h] 
\centering         
\includegraphics[width=12.5cm, angle=0]{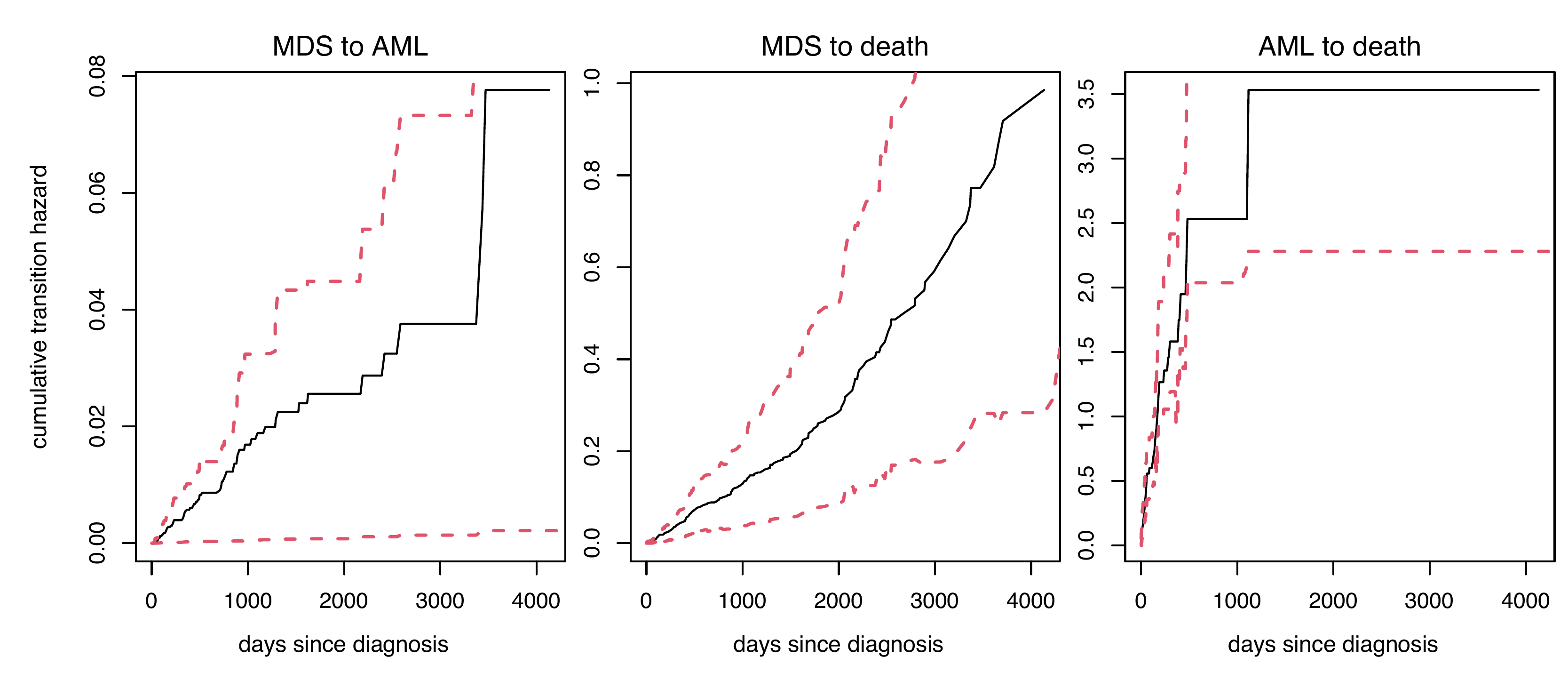} % width changes size
%\vspace*{0.25cm}     % manual adjustment of vertical spacing
\caption{Point estimates of cumulative transition hazards for a sample patient with MDS (black curve), along with $95\%$ non-parametric confidence intervals (dashed red lines).}      
\label{fig:patient1_cumhaz} % label for the figure
\end{figure}

Figure \ref{fig:patient1_cumhaz} shows three plots of estimated cumulative transition hazards for the sampled patient, one for each transition in the model, along with $95\%$ non-parametric bootstrap confidence intervals (computed with \texttt{ebmstate::boot\_ebmstate}). Throughout the plotted period, the `slope' of the cumulative hazard (i.e., the hazard rate) for the MDS to AML transition is lower than the one for the MDS to death transition, and this in turn is lower than the one for the AML to death transition. It should be recalled that the cumulative hazard estimate  is strictly non-parametric for this last transition, i.e., it is the same for all patients. The central plot of figure \ref{fig:patient1_cumhaz} suggests that, as time since diagnosis goes by, the hazard of dying in MDS increases (possibly an effect of age). On the other hand,  the hazard of dying in AML seems to decrease (slightly) with time (rightmost plot). Conclusions regarding the evolution of the AML hazard are hard to draw, since the confidence intervals for the corresponding cumulative hazard curve are very wide (leftmost plot). 

If an object generated by \texttt{msfit\_generic} is fed to \texttt{plot},  and the package \texttt{mstate} is loaded, the method \texttt{mstate:::plot.msfit} will be called. This is an efficient way of automatically plotting the cumulative hazard estimates for all transitions, but confidence interval lines (separately estimated) cannot be added.

\subsection{Computing state occupation probability estimates}
\label{sec:computing_transition_probs}
\begin{figure}[h] 
\centering         
\includegraphics[width=10.5cm, angle=0]{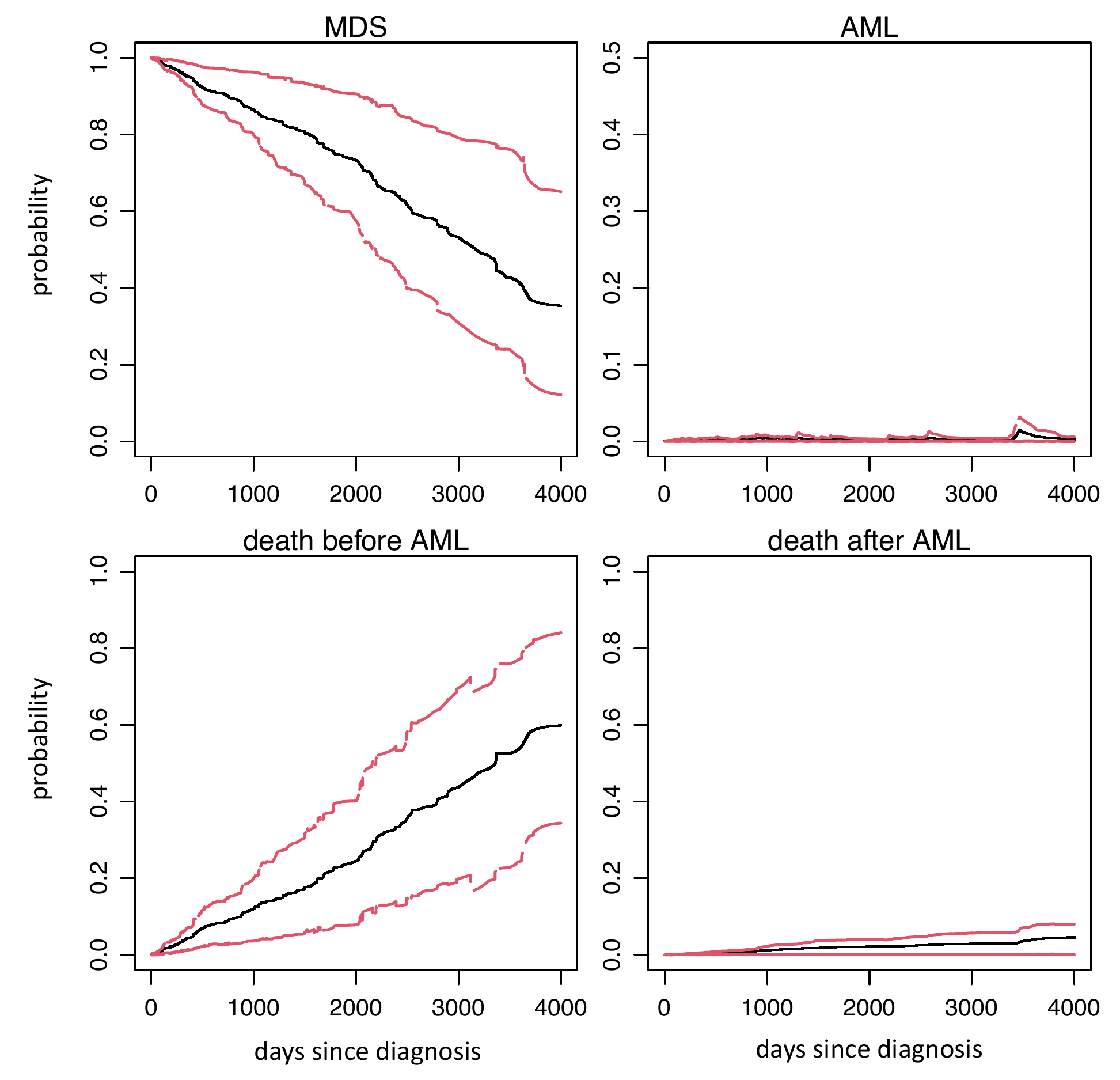} % width changes size
%\vspace*{0.25cm}     % manual adjustment of vertical spacing
\caption{Point estimates of state occupation probabilities for a sample patient with MDS (black curve), along with $95\%$ non-parametric confidence intervals (dashed red lines).}      
\label{fig:patient1_transProbs} % label for the figure
\end{figure}

The functions \texttt{probtrans\_mstate}, \texttt{probtrans\_ebmstate} and \texttt{probtrans\_fft} compute estimates of state occupation probabilities for a given \texttt{msfit} object.
All three functions generate objects of class \texttt{probtrans} that can be fed to the  \texttt{plot.probtrans} method from the package \texttt{mstate}.
The first of these functions should only be used for clock-forward models, as it relies on product-limit calculations. It calls the method \texttt{probtrans\_mstate.default}, if the \texttt{msfit} object was generated by \texttt{msfit\_generic.default}, or the method \texttt{probtrans\_mstate.coxrfx}, if it was generated by \texttt{msfit\_generic.coxrfx}. Both methods are identical to the function \texttt{probtrans} in the \texttt{mstate} package, with the reserve that \texttt{probtrans\_mstate.coxrfx}  does not allow the computation of the variances or covariances of the state occupation probability estimator. 
The functions \texttt{probtrans\_ebmstate} and \texttt{probtrans\_fft} are the functions in \texttt{ebmstate} for computation of state occupation probability estimates under clock-reset models with a tree-like transition structure.  When using \texttt{probtrans\_fft} (the faster of these two functions), three arguments must be supplied: the initial state of the process whose state occupation probabilities one wishes to compute, the \texttt{msfit} object, and a vector of positive and increasing time points starting from and including zero.   This last argument, denoted \texttt{time}, is crucial for precision: the density of time points and the upper time limit should be increased until the estimated curves become stable.  The following line of code computes point estimates of state occupation probabilities for the sample patient. 
\small
\begin{lstlisting}
probtrans_object<-probtrans_fft("MDS",msfit_object, time)
\end{lstlisting}
\normalsize
Estimates are shown in figure \ref{fig:patient1_transProbs}, along with $95\%$ non-parametric, bootstrap confidence intervals (computed using \texttt{ebmstate::boot\_ebmstate}). For this particular patient, the estimated probability of being dead after AML remains below 0.1 throughout a period of 10 years from the MDS diagnosis; if the patient does reach AML, death is expected to happen quickly thereafter, as reflected in the very low estimates for the probability of being in AML at any point in time.

\begin{figure}[h] 
\centering         
\includegraphics[width=12.5cm, angle=0]{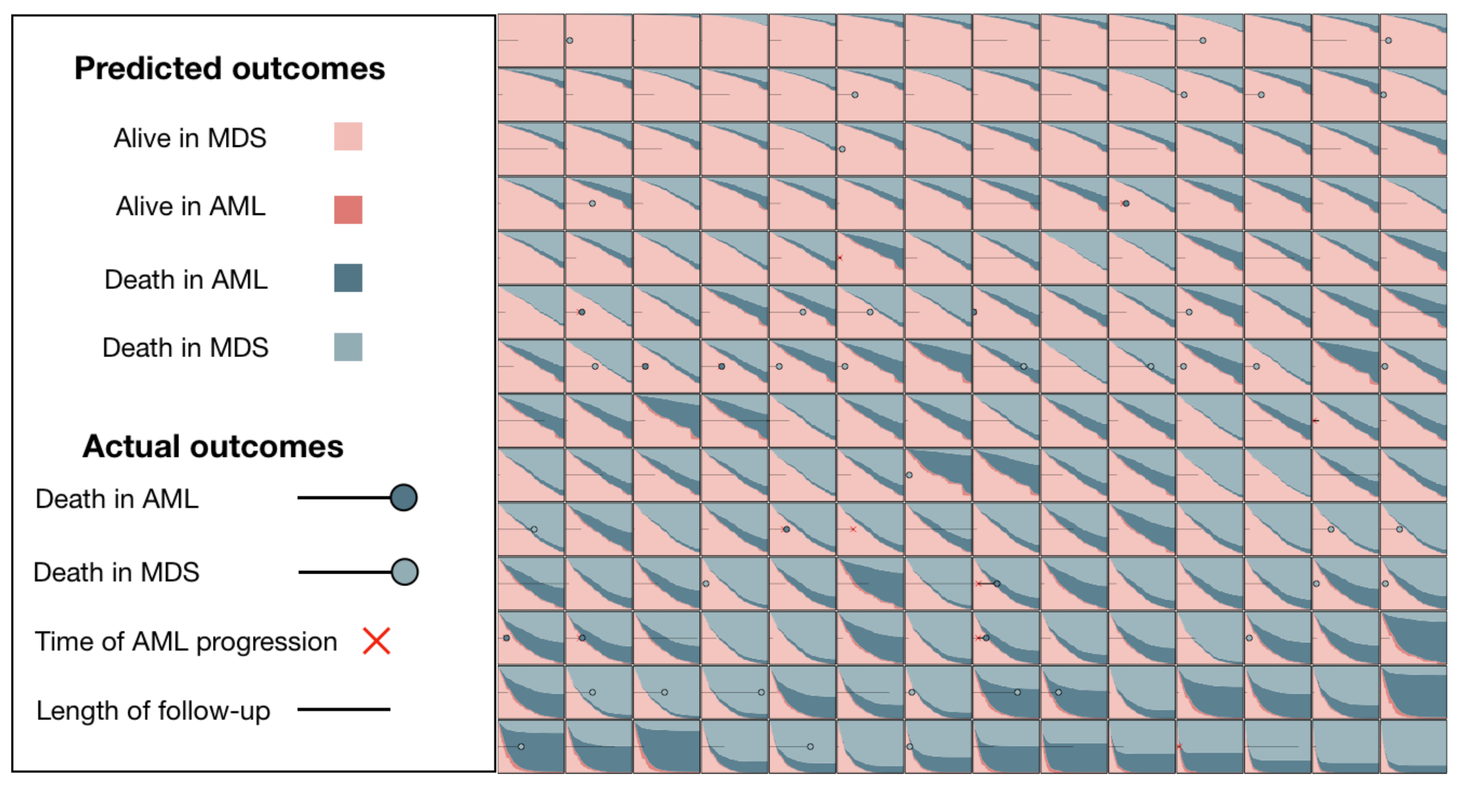} % width changes size
\vspace*{-0.25cm}     % manual adjustment of vertical spacing
\caption{Leave-one-out estimates of state occupation probabilities for a random sample of 196 individuals with MDS. The $x$-axis measures time since diagnosis and runs from 0 to 10 years.}      
\label{fig:trans_probs_mosaic} % label for the figure
\end{figure}

The plots of figure \ref{fig:trans_probs_mosaic} show leave-one-out personalised estimates of disease progression for a random sample of 196 patients, ordered by overall survival probability at 5 years since diagnosis. 
As shown in the following block of code, these were built using \texttt{ebmstate::loo\_ebmstate}. 
\small
\begin{lstlisting}
# Some arguments for the function loo_ebmstate: 
patient_IDs<-sample(unique(mstate.data$id),14*14)
mstate.data.expanded<-mstate::expand.covs(mstate.data,
	covs = names(mstate.data)[!names(mstate.data%in%c("id","from","to"
	,"trans","Tstart","Tstop","time","status","strata")],append = T)

# Leave-one-out outcome predictions:
loo_object<-loo_ebmstate(mstate.data,mstate.data.expanded,
	groups,patient_IDs,initial_state="MDS",tmat,
	time_model="clock-reset")
\end{lstlisting}
\normalsize
The mosaic plot shows a wide range of prognoses. Quite importantly, a substantial number of patients have a very low estimated probability of developing AML during the first 10 years following diagnosis, and should perhaps be spared the burden of aggressive treaments. The validity of such conclusions requires however that the model fares well in terms of prediction accuracy. Here further investigations would be needed.

\section{DISCUSSION}

We have shown that \texttt{ebmstate} is suitable for higher-dimensional, multi-state survival analysis, and that it is both efficient and easy-to-use.  To a significant extent, the user-friendliness of \texttt{ebmstate} stems from the fact that it was not built `from the ground up'.  Instead, we produced a package that is more easily accessible to the many users of \texttt{mstate} by taking advantage of whichever features of this package were useful to our method and by eliminating redundancies. 
The connection between \texttt{ebmstate} and \texttt{mstate} is based on the fact that the function \texttt{CoxRFX} takes the same type of input and produces the same type of output as \texttt{coxph} from the package \texttt{survival}, and the function \texttt{probtrans\_fft} (or \texttt{probtrans\_ebmstate}) has the same type of input and output as \texttt{probtrans} from \texttt{mstate} (as shown in figure \ref{fig:workflow}).

We also sought to improve our package's user-friendliness by making it as efficient as possible.  The reduction of computational cost is based on two features. First, our empirical Bayes method relies on an expectation-maximisation algorithm that estimates both the parameters and the hyper-parameters of the model, i.e., no further tuning of the model is required. Second, in \texttt{ebmstate}, the computation of state occupation probability estimates relies on analytical results rather than on simulation: not only for clock-forward models, where we import from \texttt{mstate} a product-limit estimator,  but also for clock-reset models, where we implement our own estimator based on a convolution argument and the fast Fourier transform.

To our knowledge, \texttt{ebmstate} is the first R package to put together a  framework for multi-state model estimation that is complete and suitable for higher-dimensional data. 
 It does so by implementing point and interval estimators of regression coefficients, cumulative transition hazards and state occupation probabilities, under regularised multi-state Cox models.  
 In section \ref{sec:estimator_performance}, the results of the simulation study suggest that for data sets with 100 patients or more and a ratio of $p$ (patients) to $n$ (coefficients per transition) greater than 0.1, the standard Cox model estimator is clearly outperformed by the empirical Bayes one when it comes to the estimation of relative hazards and state occupation probabilities of an out-of-sample patient, or the regression coefficients of the model. However, the same study suggests that using an empirical Bayes method instead of a fully non-parametric one is of limited or no value in settings where $p/n \geq 1$. This loss of usefulness can already happen for $p/n\leq 1/2$ when it comes to the estimation of the relative hazards of an out-of-sample patient, especially for transition structures with multiple competing transitions.

 As mentioned in previous sections, \texttt{ebmstate} imports a product-limit estimator from \texttt{mstate} that targets the state occupation probabilities of patients with \textit{time-fixed} covariate vectors. However, these estimators are extendible to models with time-dependent covariates, as long as these are external and the estimates are conditional on specific covariate paths \citep[][p. 142]{Aalen2008}. For piecewise constant covariates, it is likely that such an adaptation could be obtained by combining transition probability estimates obtained for each period in which the covariates are fixed. While no significant theoretical obstacles are foreseen in this matter, the computer implementation for more than a single piecewise constant covariate is likely to be a laborious task. We have left it therefore for future work.
 
\section*{Supplementary Materials}
The file `ESM\_1.html' contains additional simulation results and theoretical demonstrations.   Additional details on the analysis of the MDS data set are given in the file `ESM\_2.html'.
 
 \section*{Acknowledgements}
The authors are supported by grant NNF17OC0027594 from the Novo Nordisk Foundation.

\clearpage
\renewcommand{\refname}{{\LARGE References}}
{\normalsize
\bibliographystyle{Chicago}
\bibliography{manuscript}}
\clearpage

%{\LARGE \textbf{Figures}}
%\vspace{3cm}

\enlargethispage{3.0\baselineskip}

\end{document}